\documentclass[aps,prl,twocolumn]{revtex4}
\usepackage{amssymb}
\usepackage{epsfig}
\usepackage{graphicx}
\usepackage{amsmath}
\usepackage{amsfonts}

\begin{document}

\title{Quantum probe and design for a chemical compass with magnetic nanostructures}
\author{Jianming Cai}
\affiliation{Institut f\"ur Quantenoptik und Quanteninformation
der \"Osterreichischen Akademie der Wissenschaften, Innsbruck, Austria}

\date{\today}

\begin{abstract}
Magnetic fields as weak as Earth's may affect the outcome of certain photochemical reactions that go through a radical pair intermediate. When the reaction environment is anisotropic, this phenomenon can form the basis of a chemical compass and has been proposed as a mechanism for animal magnetoreception. Here, we demonstrate how to optimize the design of a chemical compass with a much better directional sensitivity simply by a gradient field, e.g. from a magnetic nanostructure. We propose an experimental test of these predictions, and suggest design principles for a hybrid metallic-organic chemical compass. In addition to the practical interest in designing a biomimetic weak magnetic field sensor, our result shows that gradient fields can server as powerful tools to probe spin correlations in radical pair reactions.
\end{abstract}

\maketitle

{\it Introduction.---} Recently, there has been increasing interest in quantum biology namely investigating quantum effects in chemical and biological systems, e.g., light harvesting systems \cite{LHC}, avian compass \cite{Cai10prl,Ved09,Kom09,JonHor10} and olfactory sense \cite{SME}. The main motivation is to understand how quantum coherence (entanglement) may be exploited for the accomplishment of biological functions. As a key step towards this goal, it is desirable to find tools that can detect quantum effects under ambient conditions. The ultimate goal of practical interest in studying quantum biology is to learn from nature and design highly efficient devices that can mimic biological systems in order to complete important tasks, e.g. collecting solar energy and detecting weak magnetic field.

As an example of quantum biology, the radical pair mechanism is an intriguing hypothesis \cite{Sch78} to explain the ability of some species to respond to weak magnetic fields \cite{WilRev,JohRev,HorRev}, e.g. birds \cite{Ritz00,Wil01,Ritz04}, fruit flies \cite{Geg08,Geg10}, and plants \cite{Ahm06}. A magnetochemical compass could find applications in remote magnetometry, in magnetic mapping of microscopic or topographically complex materials, and in imaging through scattering media \cite{Yang09}. It was demonstrated that a synthetic donor-bridge-acceptor compass composed of a linked carotenoid (C), porphyrin (P), and fullerene (F) \cite{Kim08} can work at low temperature (193 K). It is surprising that such a triad molecule is the only known example that has been experimentally demonstrated to be sensitive to the geomagnetic field (yet not at room temperature). It is currently not known how one might construct a biomimetic or synthetic chemical compass that functions at ambient temperature.

\begin{figure}[b]
\begin{center}
\includegraphics[width=8.8cm]{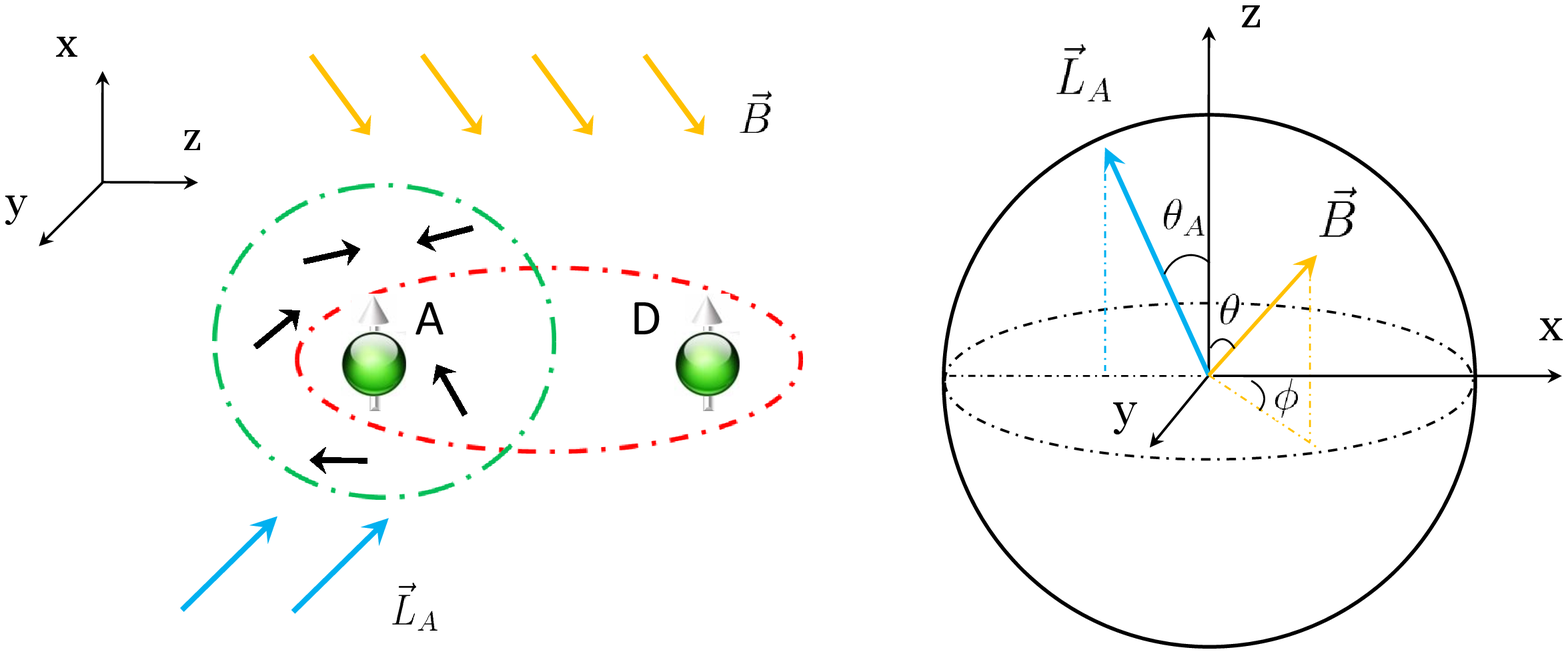}
\end{center}
\caption{(Color online) Left: A radical pair, coupled with the surrounding nuclear spins (black arrow), in a weak magnetic field $\vec{B}$ to be measured (yellow) and a strong magnetic gradient $\vec{L}_{A}$ (blue), due to e.g. a magnetic nanostructure. The outcome of a reaction depends on the direction of the weak field $\vec{B}$. Right: The directions of $\vec{B}$ and the gradient field at the location of the acceptor $\vec{L}_{A}$ depicted in the molecular coordinate frame.}\label{setup}
\end{figure}

In this Letter, we approach to the goals of studying quantum biology in the context of chemical compass by demonstrating that a suitably designed gradient field can significantly improve the performance of a model chemical compass (apart from increasing the intersystem crossing rate \cite{Coh09}), see Fig.~\ref{setup}. It also opens a possible route to probe spin correlations of radical pairs and thereby investigate the role of quantum effects in spin chemistry. The gradient field is strong at the location of one spin, and approximately zero at the other. Such a field can be created in the vicinity of a hard ferromagnetic nanostructure \cite{Coh09}, by applying a spatially uniform bias field that cancels the field of the nanostructure in a small region of space. In essence, the strong gradient field at one spin can substitute for strong anisotropic hyperfine couplings required for a purely molecular compass. This geometry provides a more significant anisotropy and thereby shows much larger directional sensitivity than does the conventional compass mechanism based only on anisotropic hyperfine couplings. Without requiring extra nuclear spins, the present model can work merely with two electron spins and thereby much simplifies quantum simulation of a chemical compass with, e.g. quantum dots and Nitrogen-vacancy centers in diamond. With more freedom to tune parameters in a better controllable environment, such kind of quantum simulations would be very helpful to understand the recombination process of radical pairs, in particular, whether and how quantum measurement and Zeno effect take place \cite{Kom09,JonHor10}.

{\it Chemical compass mechanism.---} Many chemical processes involve a radical pair intermediate, in which each radical has an unpaired electron coupled to an external magnetic field and a few nuclei via the Hamiltonian \cite{Ste89}
\begin{equation}
H=\sum_{k=A,D}H_{k}=-\gamma _{e}
\sum_{k}\vec{B}_{k}\cdot \vec{S}_{k}+\sum_{k,j} \vec{S}_{k}\cdot \hat{\lambda}
_{k_{j}}\cdot \vec{I}_{k_{j}}  \label{Hamil}
\end{equation}
where $\gamma _{e}=-g_{e}\mu _{B}$ is the electron gyromagnetic
ratio, $\hat{\lambda}_{k_{j}}$ denote the hyperfine coupling tensors
and $\vec{S}_{k}$, $\vec{I}_{k_{j}}$ are the electron and nuclear
spin operators respectively. In our model, the magnetic field consists of two parts: $\vec{B}_{k}=\vec{B}+\vec{L}_{k}$, where the directional information about $\vec{B}$ is what one wants to infer from the radical pair reaction, and $\vec{L}_{k}$ is the local gradient field applied to each radical and is independent of $\vec{B}$. The spin relaxation and decoherence times resulting from the factors other than hyperfine interactions are assumed to be considerably longer than the radical pair lifetime \cite{Ritz00,Ved09}, to maximize sensitivity to weak magnetic fields \cite{Ritz09}. In many photochemical processes, the radical pair is created in a spin-correlated electronic singlet state $|\mathbb{S}\rangle =\frac{1}{\sqrt{2}}(\left\vert \uparrow
\downarrow \right\rangle -\left\vert \downarrow \uparrow
\right\rangle )$ within the timescale of picoseconds. The nuclear spins start at thermal equilibrium, which under ambient conditions leads to an approximate density matrix as $\rho _{n}(0)=\bigotimes_{j}
\mathbb{I}_{j}/d_{j}$, where $d_{j}$ is the dimension of the $j$th nuclear spin and $\mathbb{I}_{j}$ is the identity matrix. The Zeeman splitting from a magnetic field $\vec{B}$ as weak as the geomagnetic field is much smaller than the thermal energy at ambient temperature. Nonetheless, the field can influence the non-equilibrium electron spin dynamics and thereby determine the ratio of the chemical product from the singlet or triplet recombination as long as the thermalization time is longer than the reaction time.

In experiments, one may measure different quantities that are dependent on the weak magnetic field $\vec{B}$. Here we consider a simple first-order recombination reaction of the singlet radical pairs. We note that there is some controversy over how to describe the radical pair reactions (see e.g. \cite{Kom09,JonHor10,Maeda10,Shu10}). Nevertheless, the conventional phenomenological density matrix approach \cite{Ste89} works well in most cases, in particular when the singlet and triplet recombination rates are the same (i.e. $k_{S}=k_{T}=k$) \cite{COMP}. We adopt this method and calculate the singlet yield as $\Phi_{S}=\int_{0}^{\infty}f(t)P_{S}(t)dt$, where $f(t)=k e^{-k t}$ is the radical reencounter probability distribution, and $P_{S}(t)=\langle \mathbb{S}|\rho_{s}(t)|\mathbb{S}\rangle $ is the singlet fidelity for the electron spin state $\rho_{s}(t)$ at time $t$. The integration of $\Phi_{S}$ was performed following the method in \cite{Bro76,SI}.

{\it Gradient enhancement of magnetic field sensitivity.---} We starts from an optimally designed hyperfine compass model, one radical has strong and anisotropic hyperfine interactions, and the other radical has no hyperfine couplings \cite{Ritz09}. We arbitrarily choose to call the first radical the acceptor, A, and the second the donor, D, though nothing that follows depends on this designation. Ritz and coworkers proposed that the radical pair FADH$^{.}$-O$_{2}^{.-}$ meets this criterion, and they further speculated that this radical pair may be responsible for the magnetoreception of European robins \cite{Ritz09}, see also \cite{Sol09}. Without loss of the essential physics, we take the hyperfine couplings ($\sim $ G) from FADH$^{.}$-O$_{2}^{.-}$ \cite{Hore2003s} for our calculations.

\begin{figure}[h]
\begin{center}
\begin{minipage}{9cm}
\includegraphics[width=4.5cm]{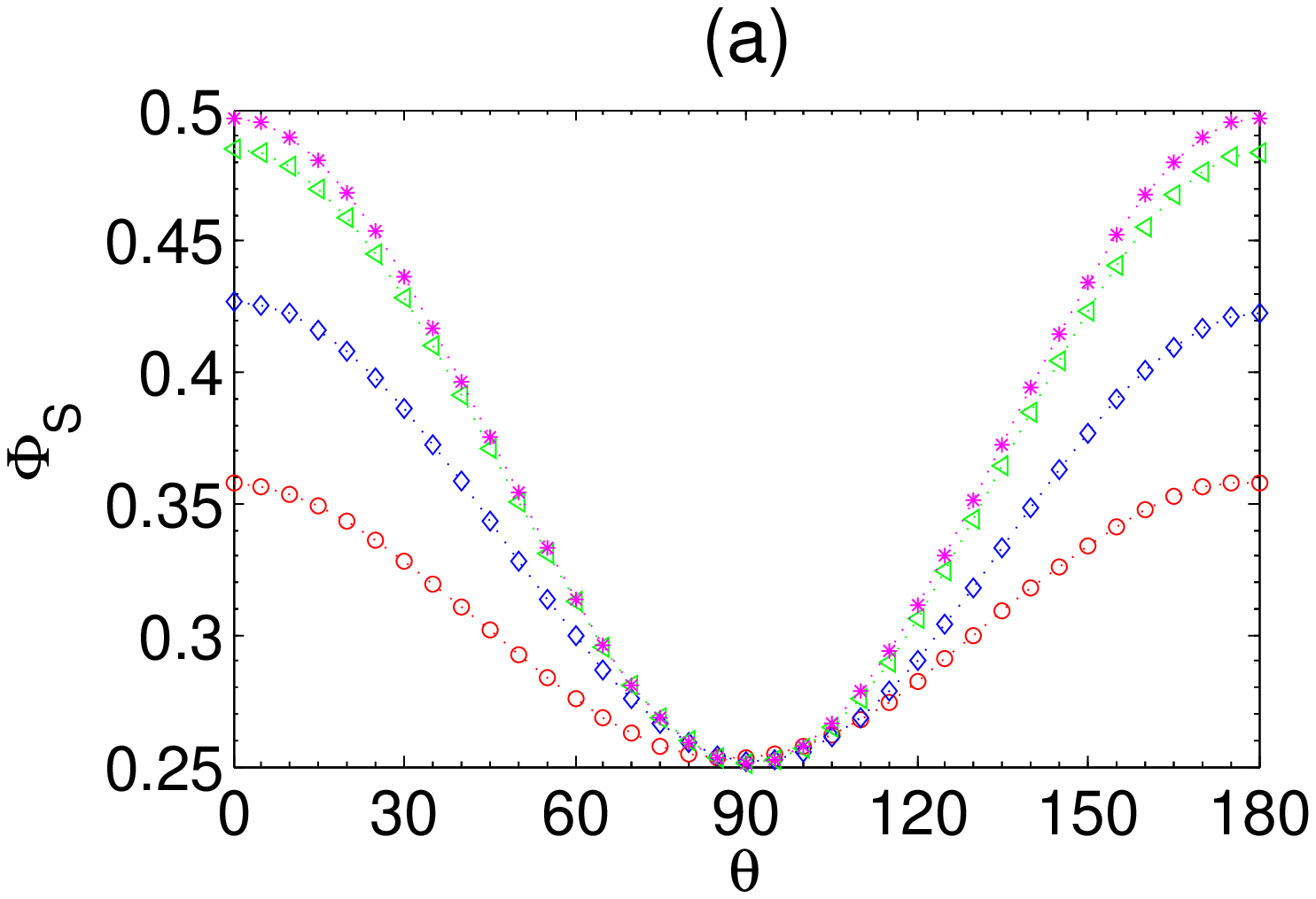}
\includegraphics[width=4.4cm]{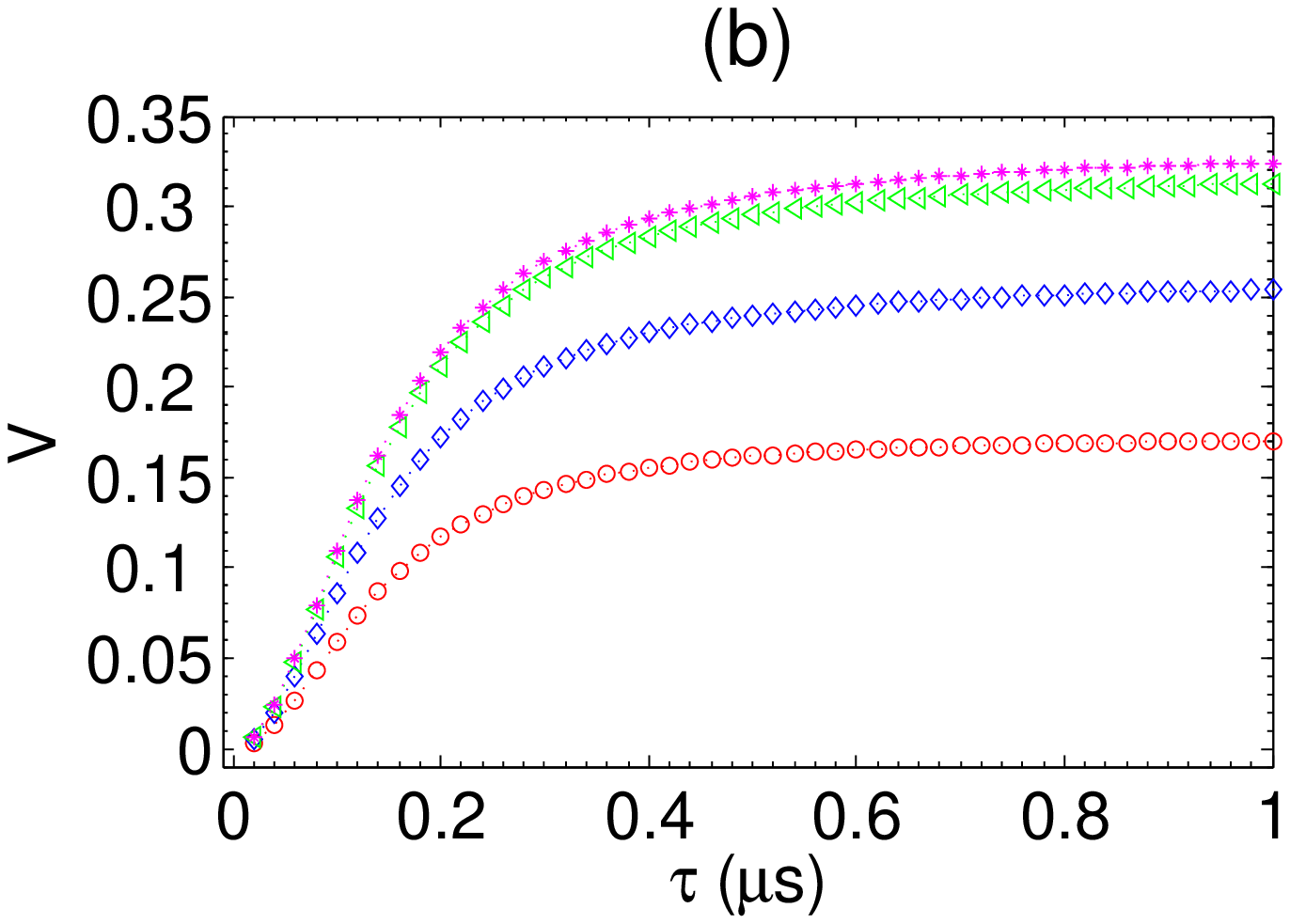}
\end{minipage}
\end{center}
\caption{(Color online) Magnetic field sensitivity of a chemical compass enhanced by a gradient field. (a) Singlet yield $\Phi_{S}$ as a function of the angle $\theta$ of the weak magnetic field $\vec{B}$ ($B=0.46$ G) with different gradient field strengths on the acceptor, i.e. $L_{A}=$ 0 G (red, $\cdots o \cdots$), 20 G (blue, $\cdots \diamond \cdots$), 40 G (green, $\cdots \triangle \cdots$), 80 G (purple, $\cdots * \cdots$), while $L_{D}=0$. The recombination rate $k=0.5 \mu s^{-1}$. (b) Visibility $V$ as a function of the radical pair lifetime $\tau=1/k$. The direction of the gradient field $\vec{L}_{A}$ is set as $\theta_{A}=0$. The same values of $L_{A}$ are used as in (a).}\label{SinYVF}
\end{figure}

We define the molecular frame as the coordinate system, and the weak magnetic field $\vec{B}$ can be represented as $\vec{B}=B(\sin\theta \cos\phi$, $\sin\theta \sin\phi, \cos \theta)$. The gradient field induces different local fields on two radicals. We assume that the gradient field on the acceptor radical is $\vec{L}_{A}=L_{A}(\sin\theta_{A}, 0, \cos \theta_{A})$ while $\vec{L}_{D}=0$ for the donor radical. The strength of the weak magnetic field to be detected is the same as the geomagnetic field, i.e. $B=0.46$ G. To demonstrate the basic idea, we first consider $\phi=0$, and then generalize to arbitrary $\phi$.

In Fig.~\ref{SinYVF} (a), we plot the singlet yield as a function of the angle $\theta$ of the weak magnetic field $\vec{B}$ with different gradient field strengths $L_{A}=$0G, 20 G, 40 G, 80 G on the acceptor. In the case of $L_{A}=0$, the directionality comes only from hyperfine anisotropy. The gradient field clearly enhances the amplitude of the direction-dependent component of the magnetic field effect (MFE). To quantify the directional sensitivity, we use the magnetic visibility defined as \cite{Cai10prl}
\begin{equation}
V=(\max \Phi_{S}-\min \Phi_{S})/(\max \Phi_{S}+\min \Phi_{S}).
\end{equation}
As the gradient field becomes larger, the sensitivity will increase and approach to a saturate best value. Fig.~\ref{SinYVF} (b) shows that for long radical pair lifetimes, the visibility with the gradient field $L_{A}=$40 G is almost twice the visibility without the gradient field. Usually, the radical pair lifetime should be very long (microseconds) to maximize the effect of weak magnetic field \cite{Ritz09}, and hence performance, of the chemical compass [Fig.~\ref{SinYVF} (b)]. This requirement places a severe constraint on the chemistry; in typical radical pair reactions the lifetime is less than 100 ns \cite{Ste89}. By increasing the overall magnitude of the visibility, gradient-enhancement broadens the range of candidate reactions for a chemical compass.

\begin{figure}[tb]
\begin{center}
\begin{minipage}{9cm}
\hspace{-0.2cm}
\includegraphics[width=4.4cm]{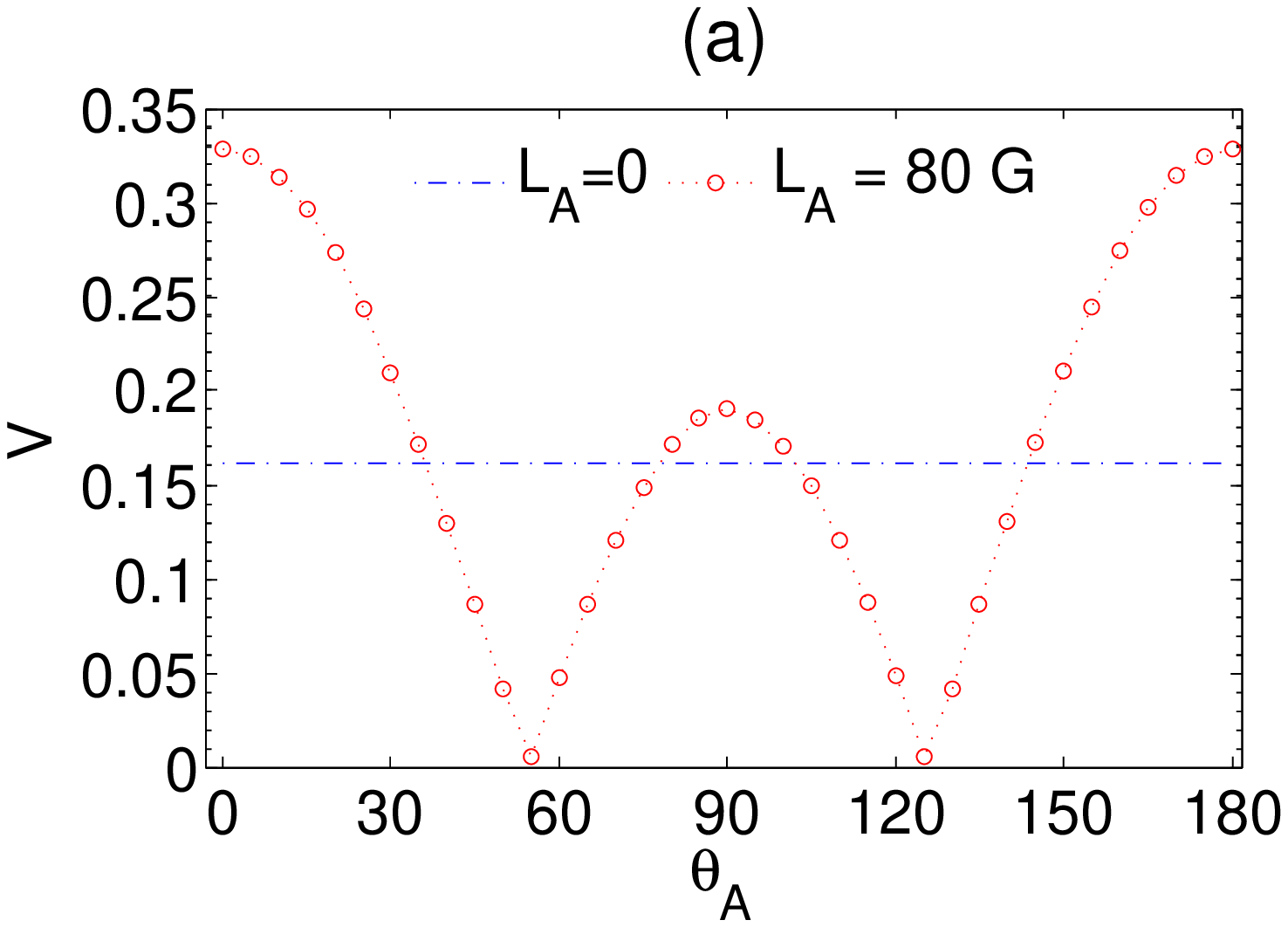}
\hspace{-0.4cm}
\includegraphics[width=4.4cm]{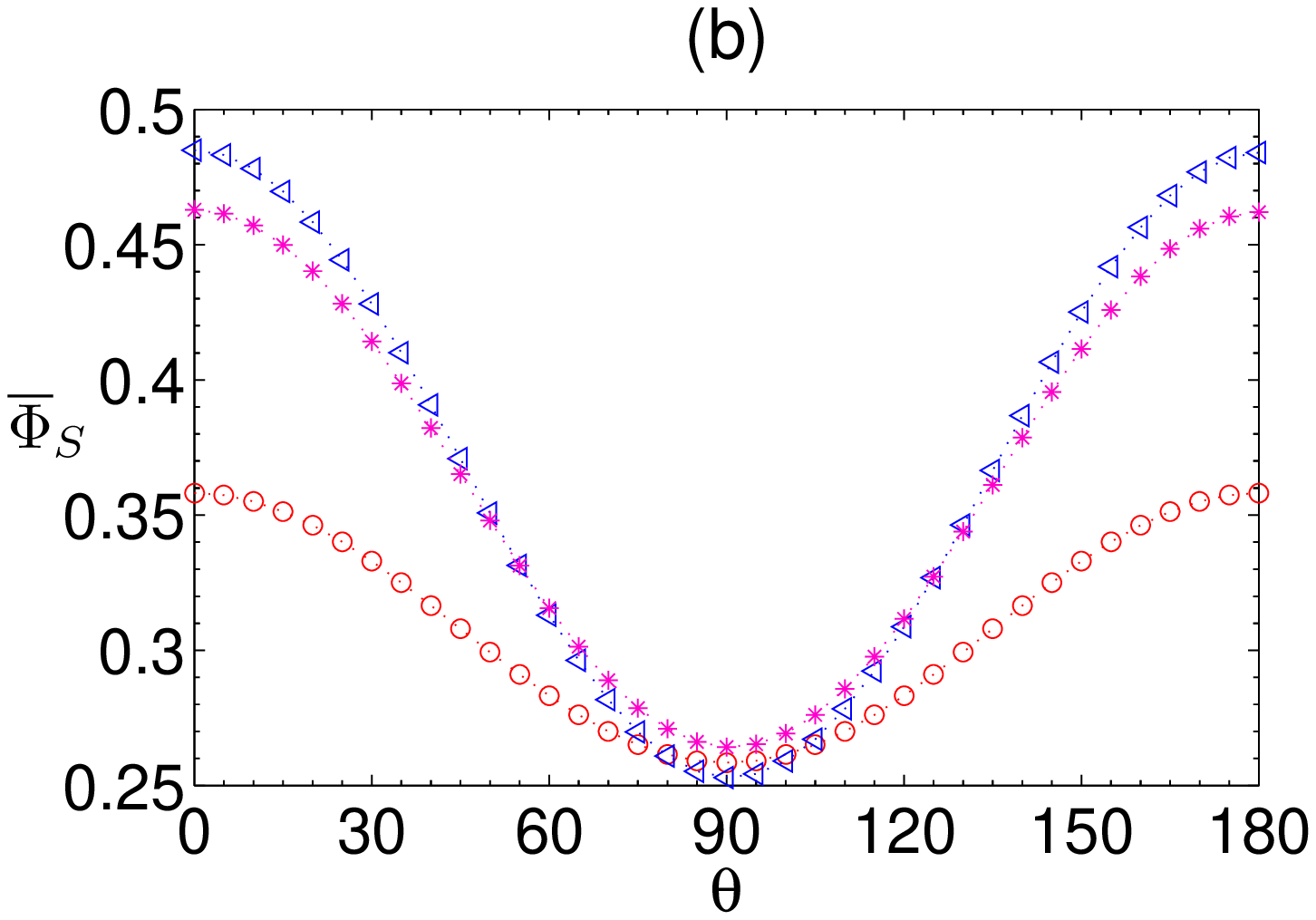}
\end{minipage}
\end{center}
\caption{(Color online) Magnetic field sensitivity in a liquid crystal experiment. (a). Visibility of the average singlet
yield $\left\langle\Phi_{S}(\theta)\right\rangle$ as a function of the angle $\theta_{A}$ of the gradient field $L_{A}=80$ G.
The blue dashed curve represents the visibility without the gradient field. (b) Ensemble average of the singlet
yield $\overline{\Phi}_{S}(\theta)$ in Eq.~(\ref{EavSinY}) from a Monte Carlo simulation of $2 \times 10^{4}$ samples
as a function of the angle $\theta$ of the weak magnetic field $\vec{B}$. The gradient field is $L_{A}=40$ G with $\theta_{A}=$0 (purple, $\cdots * \cdots$), while $L_{D}=0$.
The fluctuations of the local fields $L_{D}$ and $L_{A}$ are characterized by the 3-dimensional Gaussian distributions with the variance $\sigma_{A}=2$ G and $\sigma_{D}=0.1$ G. For comparison, we plot the singlet yield with no gradient field (red, $\cdots o \cdots$), and the one with the gradient field $L_{A}=40$ G ($\theta_{A}=$0) without fluctuations (blue, $\cdots \triangle \cdots$). In both panels the radical pair lifetime is $2 \mu s$ and $B=0.46$ G.}\label{aVSBSin}
\end{figure}

{\it Liquid crystal experiment.---} In a uniaxially oriented sample, the MFE is averaged over all values of the angle $\phi$. Such a sample is prepared by, for instance, freezing the molecules in a nematic liquid crystal in the presence of a strong magnetic field \cite{Kim08}. The ensemble-averaged MFE depends on $\theta$ only and is characterized by
\begin{equation}
\left\langle \Phi_{S} (\theta) \right \rangle=\frac{1}{2 \pi} \int_{0}^{2 \pi} \Phi_{S}(\theta,\phi)d\phi
\label{aSinY}
\end{equation}
It can be seen from Fig.~\ref{aVSBSin} (a) that the enhancement of the sensitivity can still be observed with the average signal $\left\langle\Phi_{S}(\theta) \right\rangle$ by choosing appropriate values of $\theta_{A}$.

To induce the gradient field as above, one feasible way is to use magnetic nanostructures \cite{Coh09}.
We model the nanocrystal as a uniformly magnetized sphere, in which case the external magnetic field is the same as that of a point dipole of magnetic moment $\textbf{m}$ located at the center of the sphere \cite{dip99}. We denote the position relative to the center of the sphere by the vector $\textbf{r}$, and assume that both $\textbf{r}$ and $\textbf{m}$ lie along the z-axis. The magnetic field at $\textbf{r}$ is
\begin{equation}
\textbf{B}(\textbf{r})=\frac{\mu_{0} m}{2\pi r^3} \hat{\textbf{r}},
\end{equation}
where $\mu_{0}=4\pi\times 10^{-7}$N$\cdot$A$^{-2}$ is the permeability of free space, the magnetic moment $m=M\varrho \Omega$ with $M$ the specific magnetization, $\rho$ is the material density, $\Omega=\frac{4}{3}\pi R^3$ is the volume of the particle and $R$ is its radius. The parameters for the typical magnetic material F${\mbox{e}_3}$O$_{4}$ are $M=43$ A$\cdot$m$^2\cdot$ kg$^{-1}$, $\varrho=5210$ kg$\cdot$m$^{-3}$ \cite{Coh09}. For molecules with a separation $r_{AD}$ between two radicals a few nanometers \cite{Val05}, it is sufficient for a nanoparticle to induce a large local field imbalance ($\sim 10$ G) on the donor and acceptor. For example, using a F${\mbox{e}_3}$O$_{4}$ nanoparticle with the radius $R=15$ nm, it is possible to induce the local field difference as large as $\sim 40$ G between the position $r_{A}=35$ nm and $r_{D}=r_{A}+r_{AD}= 38.5$ nm (assuming $r_{AD}=3.5$ nm). By generating an additional homogenous field to compensate the field at the position $r_{D}$, we can effectively obtain the gradient field on the donor and acceptor molecule as $L_{A}\simeq 40$ G and $L_{D}= 0 $ G respectively.

To see whether the effect of the gradient field shown above can manifest with experimental imperfections, we take into account the fluctuations of $\vec{L}_{A}$ and $\vec{L}_{D}$ by modeling the fluctuation as the three-dimensional Gaussian distribution $f(\mathbf{\Delta}_{i})=\frac{1}{(2\pi \sigma_{i}^2)^{3/2}}\exp{(-\frac{|\mathbf{\Delta} _{i}|^{2}}{2\sigma_{i}^{2}})}$ ($i=A, D$) with $\sigma_{A}=2$ G and $\sigma_{D}=0.1$ G. Therefore, the ensemble average of $\left\langle \Phi_{s}(\theta) \right\rangle$ in Eq.~(\ref{aSinY}) is
\begin{equation}
\overline{\Phi}_{S}(\theta)= \int \left\langle \Phi_{S}(\theta)\right\rangle|_{\mathbf{\Delta}_{A}, \mathbf{\Delta}_{D}} f(\mathbf{\Delta}_{A})f(\mathbf{\Delta}_{D})d\mathbf{\Delta}_{A}d\mathbf{\Delta}_{D}\label{EavSinY}
\end{equation}
where $\left\langle \Phi_{S}(\theta)\right\rangle|_{\mathbf{\Delta}_{A}, \mathbf{\Delta}_{D}}$ is the average singlet yield when the local fields on the acceptor and donor molecules are $\vec{L}_{A} +\mathbf{\Delta}_{A}, \vec{L}_{D}+\mathbf{\Delta}_{D}$ respectively. We have used Monte Carlo simulations to calculate the above ensemble average in Eq.(\ref{EavSinY}). In Fig.~\ref{aVSBSin} (b), we see that the enhancement from the gradient field can still be observed.

{\it Probe spin correlations in a chemical compass.---} Besides the significant enhancement of the directional sensitivity offered by gradient fields, we now examine how they can provide new insights into the quantum dynamics of radical pair reactions. For the present model chemical compass, if the gradient field on the acceptor $\vec{L}_{A}$ dominates over the hyperfine couplings and the weak magnetic field $\vec{B}$, the singlet yield can be written as \cite{SI}
\begin{equation}
\Phi_{S} (\vec{L}_{A},\vec{B})= \frac{1}{4}-\frac{1}{4}\langle \hat{A}\otimes \hat{V} \rangle
\end{equation}
where the expectation value is calculated over the initial state, and $\hat{A}=|u_{0}\rangle\langle u_{0}|-|u_{1}\rangle \langle u_{1}|$ (with $\{|u_{0}\rangle,|u_{1}\rangle\} $ the eigen states of $\vec{L}_{A} \cdot \vec{S}_{A} $), $\hat{V}=\langle U_{D}^{\dagger} \hat{A}U_{D}\rangle$ with $U_{D}=\exp{(i \gamma_{e}t \vec{B}\cdot \vec{S}_{D} )} $ and the average taken over time weighted by $f(t)$. By choosing $\vec{L}_{A}$ in the direction of $\hat{x}$, $\hat{y}$, and $\hat{z}$, the corresponding operator $\hat{A}$ will be $\hat{X}$, $\hat{Y}$, $\hat{Z}$ respectively (which are the Pauli operators). Moreover, for each $\hat{A}$, one can choose $\hat{B}$ also in the direction of $\hat{x}$, $\hat{y}$, and $\hat{z}$ such that the operators of $\hat{V}$ (as a linear combination of Pauli operators) are linear independent, see \cite{SI}. The singlet yields corresponding to these choices of $\hat{L}_{A}$ and $\hat{B}$ lead to nine independent equations, from which we can infer the spin correlations $\langle \hat{M}\otimes \hat{N} \rangle$ for the radical pair state, where $\hat{M},\hat{N}=$ $\hat{X}$, $\hat{Y}$ or $\hat{Z}$. With these correlations, one may check whether the radical pair state violates Bell inequalities \cite{Hor96}; or obtain lower entanglement bounds of the radical pair state, see Ref. \cite{Aud06}. The above idea can be extended to monitor the dynamics of spin correlations suppose one can switch on gradient fields during the reaction.

\begin{figure}[tb]
\begin{center}
\begin{minipage}{9.0cm}
\hspace{-1.0cm}
\includegraphics[width=4.2cm]{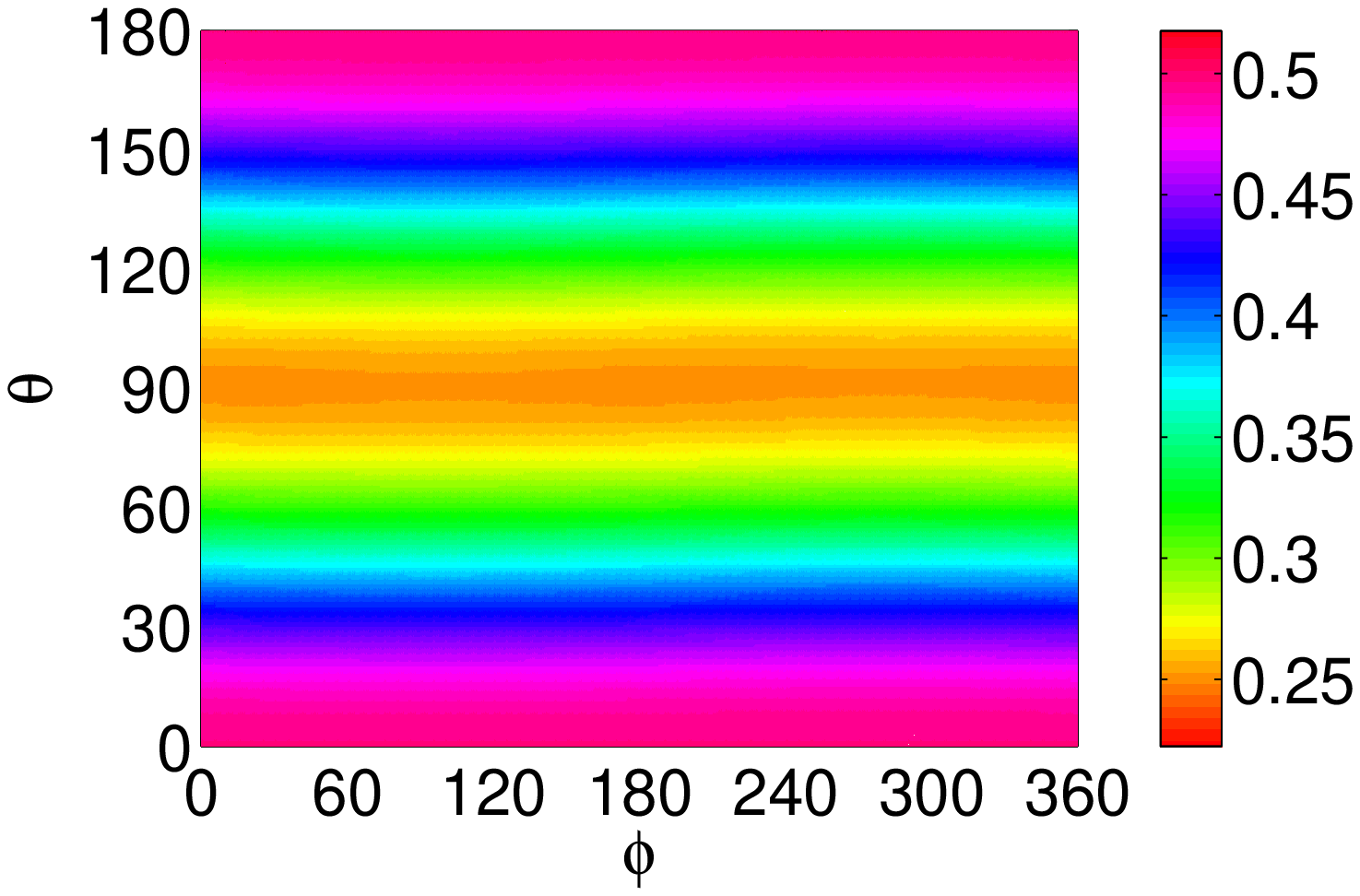}
\hspace{-0.1cm}
\includegraphics[width=1cm]{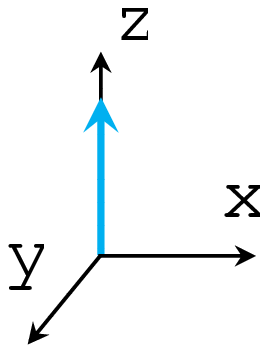}
\hspace{-0.35cm}
\vspace{0.1cm}
\includegraphics[width=4.2cm]{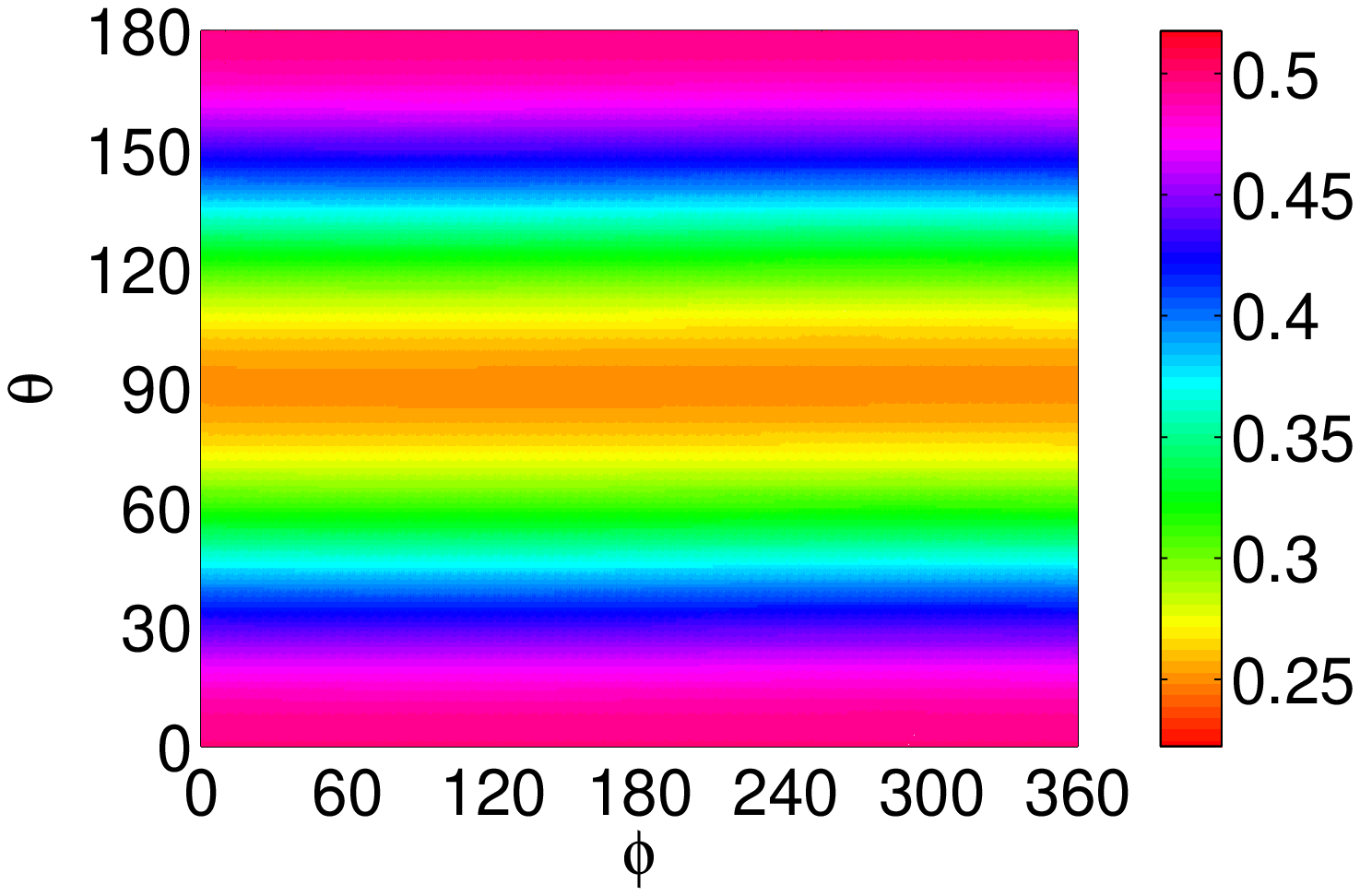}
\end{minipage}
\begin{minipage}{9.0cm}
\hspace{-1.0cm}
\includegraphics[width=4.2cm]{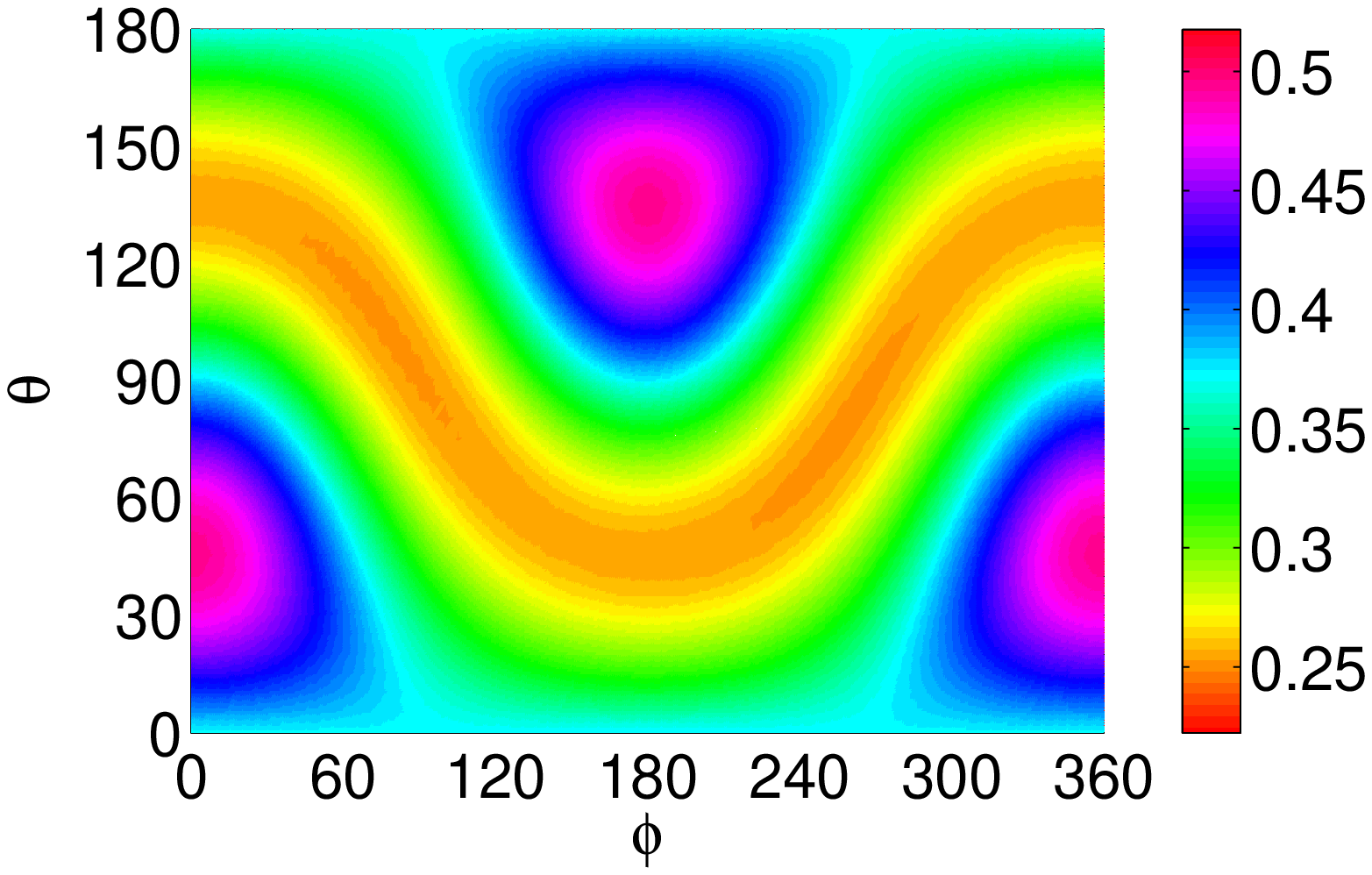}
\hspace{-0.1cm}
\includegraphics[width=1cm]{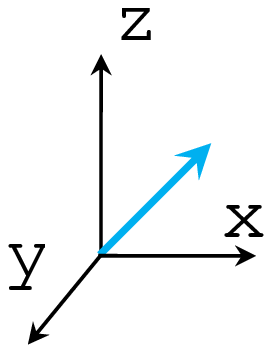}
\hspace{-0.35cm}
\vspace{0.1cm}
\includegraphics[width=4.2cm]{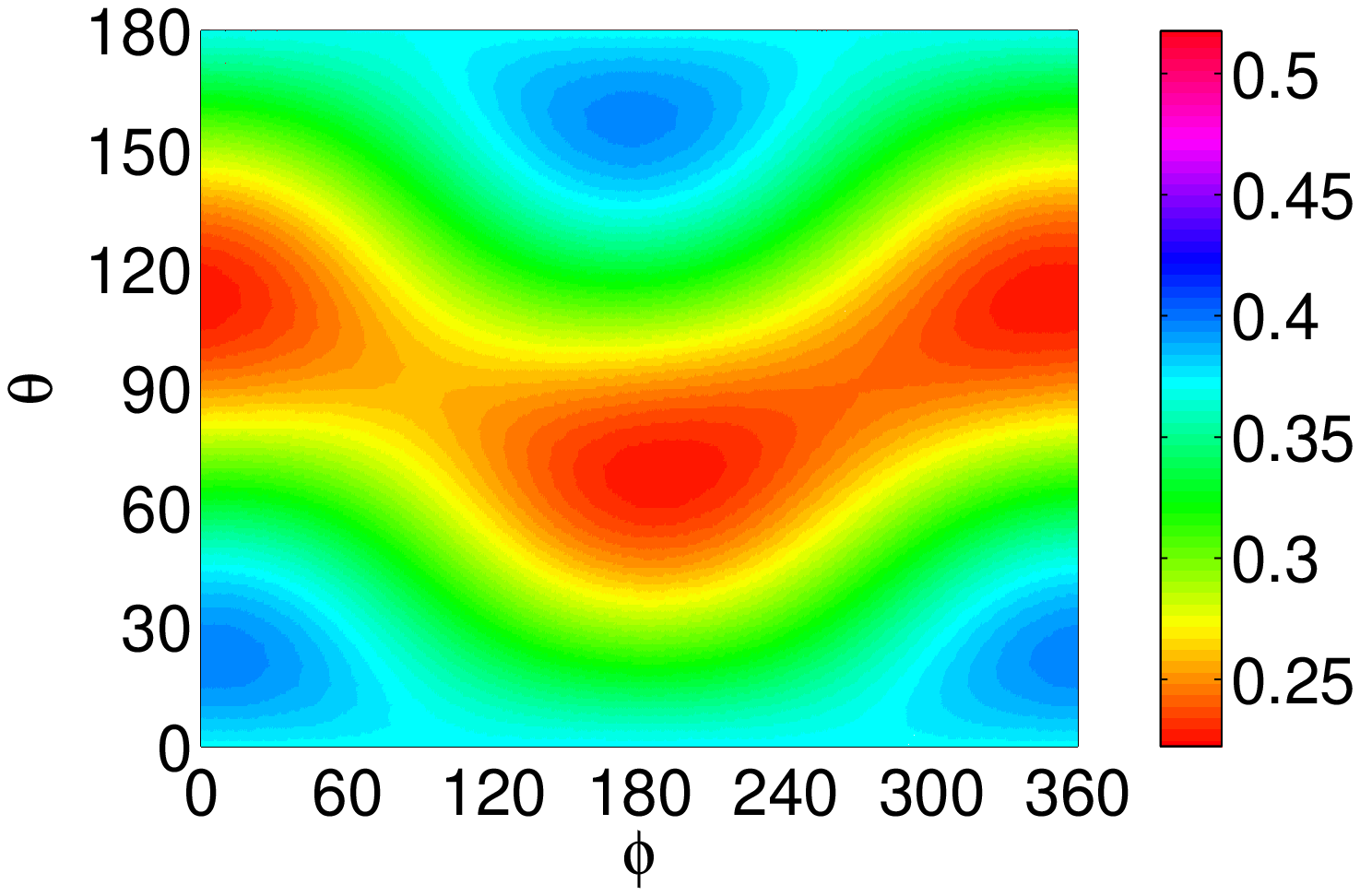}
\end{minipage}
\begin{minipage}{9.0cm}
\hspace{-1.0cm}
\includegraphics[width=4.2cm]{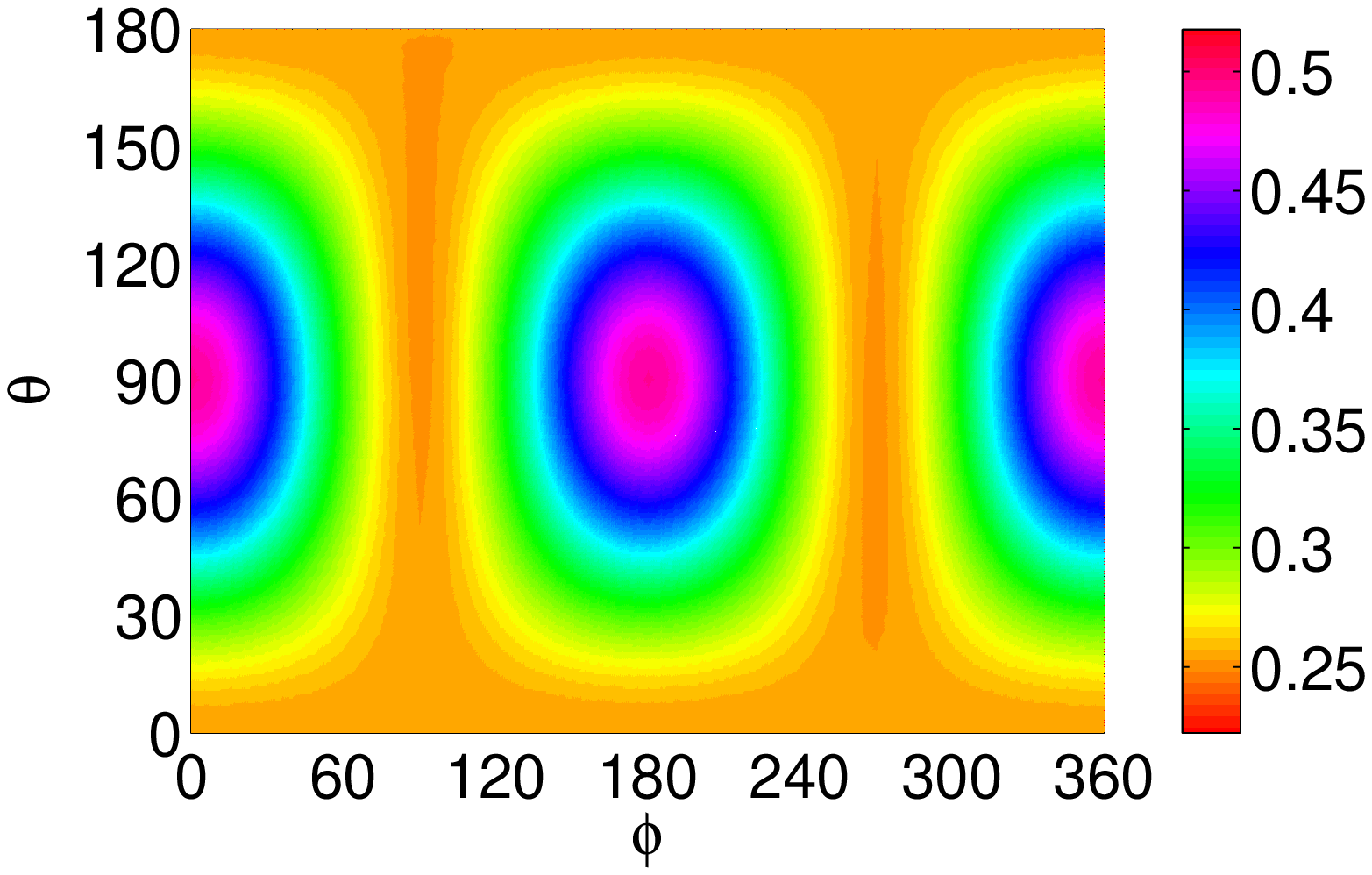}
\hspace{-0.1cm}
\includegraphics[width=1cm]{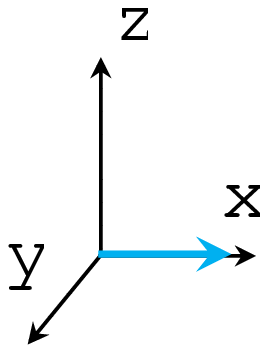}
\hspace{-0.35cm}
\includegraphics[width=4.2cm]{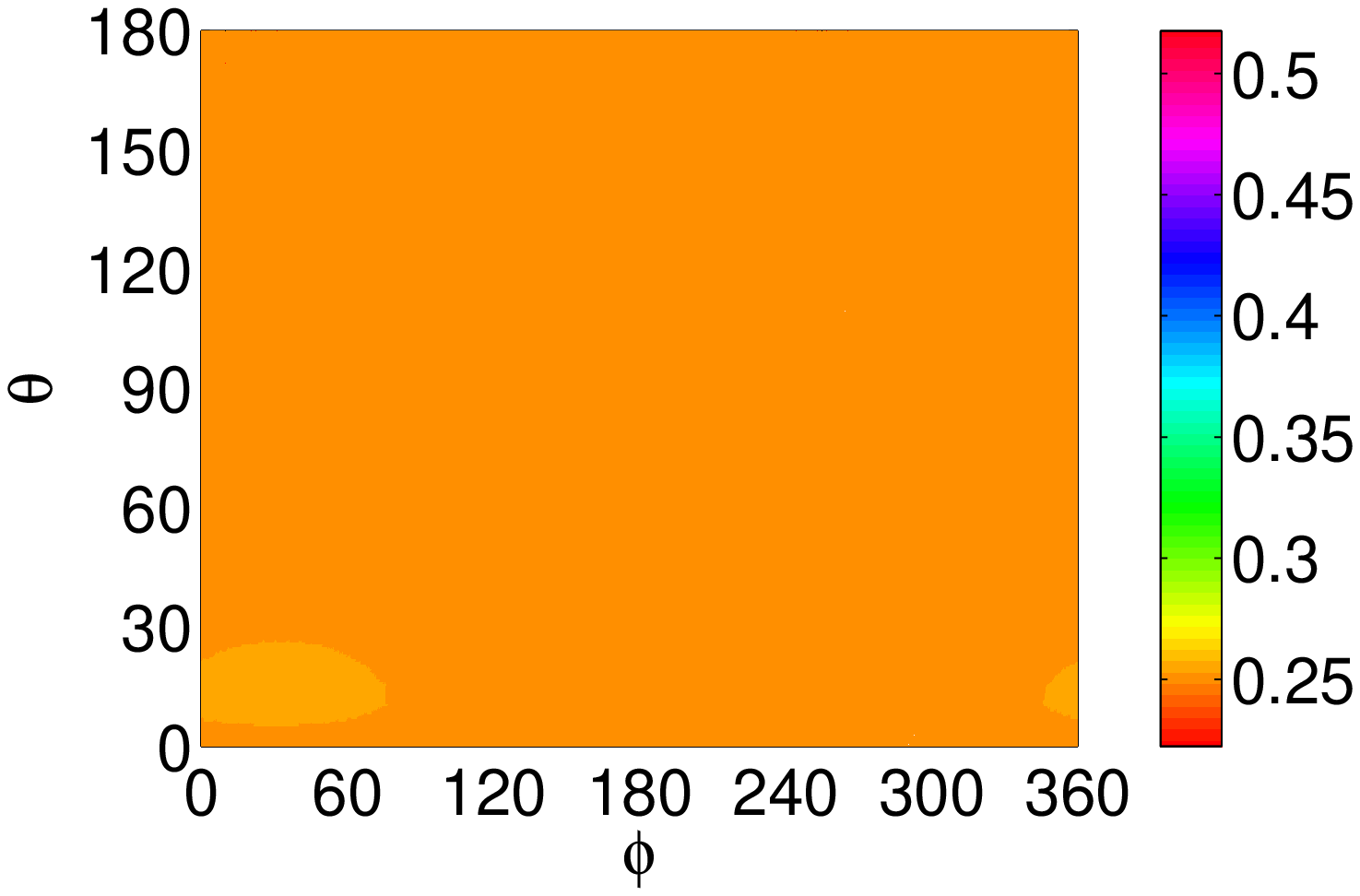}
\end{minipage}
\end{center}
\caption{(Color online) The gradient field as a tool to test the initial radical pair state. The singlet yield $\Phi_{S}$ as a function of the angles ($\theta$, $\phi$) of the weak magnetic field $\vec{B}$ with $B=0.46$ G. The gradient field on the acceptor is $L_{A}=$ 80 G, on the donor it is $L_{D}=0$. The angle of the gradient field $\theta_{A}$ with respect to the z-axis is $0$ (upper), $\frac{\pi}{4}$ (middle) and $\frac{\pi}{2}$ (lower). The patterns of the singlet yield over ($\theta$, $\phi$) from the initial singlet (left) and classically correlated state (right) are quite similar for $\theta_{A}=0$, but are very different for $\theta_{A}= \frac{\pi}{4}$ and $\frac{\pi}{2}$. The radical pair lifetime is chosen as $2 \mu s$.}\label{VTF}
\end{figure}

As an example, we show that gradient fields can distinguish the singlet and the classically correlated initial state $\rho_{c} =(\left\vert
\uparrow \downarrow \right\rangle \left\langle \uparrow \downarrow
\right\vert +\left\vert \downarrow \uparrow \right\rangle
\left\langle \downarrow \uparrow \right\vert )/2$. For systems where the radical pair lifetime is much
longer than the decoherence time, the conventional hyperfine-mediated MFE does not strongly depend on the
initial states and thus can not allow one to achieve this goal, see e.g. \cite{Cai10prl}. If the gradient
field is along the z-axis, the singlet yields are quite similar for the singlet and the classically correlated
state [see Fig.~\ref{VTF} (upper)]. However, if we vary the direction of the gradient field $\vec{L}_{A}$, then
the visibility for the singlet state will be much larger than for the classically correlated initial state [Fig.~\ref{VTF} (middle, lower)]. In particular, for the classically correlated state, the singlet yield is insensitive to the angles ($\theta$, $\phi$) for $\theta_{A}=\frac{\pi}{2}$ while for the singlet state the angular sensitivity persists [Fig.~\ref{VTF} (lower)].
The difference originates from the essential boundary between classical and quantum correlation (entanglement). The large gradient field can be viewed as a measurement of the acceptor spin along $\vec{L}_{A}$: the singlet state demonstrates perfect anti-correlation of the spins for any direction of $\vec{L}_{A}$, while for the classically correlated state this is true only in a certain direction of $\vec{L}_{A}$ (i.e. the $\hat{z}$ direction).

{\it Summary.---} We have demonstrated that a gradient field can lead to
a significant enhancement of the performance of a chemical compass. The gradient field also provides us
a powerful tool to investigate quantum dynamics of radical pair reactions in spin chemistry. In particular, it can distinguish whether the initial radical pair state is in the entangled singlet state or in the classically correlated state, even in the scenarios where such a goal could not be achieved before. These phenomena persist upon addition of partial orientational averaging and addition of realistic magnetic noise. The effects predicted here may be detectable in a hybrid system compass composed of magnetic nanoparticles and radical pairs in an oriented liquid crystalline host. Our work offers a simple method to design/simulate a biologically inspired weak magnetic field sensor based on the radical pair mechanism with a high sensitivity that may work at room temperature.

{\it Acknowledgements.---} We are in debt to Nan Yang and Adam Cohen for valuable suggestions and beneficial communications. We thank Kiminori Maeda, Gian Giacomo Guerreschi and Otfried G\"{u}hne for helpful discussions, and Hans Briegel for continuous support in this work. The work is supported by FWF (SFB FoQuS).

\onecolumngrid

\section*{Appendix}

\textsf{\bf Calculation of singlet yield.---} We adopt the method as in \cite{Bro76SI} to calculate the singlet yield. For the self-completeness, here we present a simple outline of this method. The Hamiltonian for the system (two electron spins, one of which is coupled with several surrounding nuclear spins ) is as follows
\begin{equation}
H=\sum_{k=1,2}H_{k}=-\gamma _{e}
\sum_{k}\vec{B}_{k}\cdot \vec{S}_{k}+\sum_{k,j} \vec{S}_{k}\cdot \hat{\lambda}
_{k_{j}}\cdot \vec{I}_{k_{j}}
\end{equation}
In our calculations of the main text, we have neglected the Zeeman interactions between the nuclear spins and the external magnetic field. We have included these interactions, and verified that the induced difference is very small (as the gyromagnetic ratio for a nucleus H and N is much smaller than $\gamma_{e}$). The numbers of hyperfine couplings are take from Ref.\cite{Hore2003SI}. We calculate the singlet yield as \cite{Ste89SI}
\begin{equation}
\Phi_{S}=\int_{0}^{\infty}f(t)P_{S}(t)dt \label{EXPMODEL}
\end{equation}
where $f(t)=k e^{-k t}$ is the radical re-encounter probability distribution,
and $P_{S}(t)=\langle S|\rho(t)|S\rangle $ is the singlet
fidelity for the electron spin state $\rho(t)$ at time $t$. Eq.~(\ref{EXPMODEL}) can be obtained from the conventional Haberkorn approach \cite{HabSI} in the case that the singlet and triplet recombination rates are the same, i.e. $k_{S}=k_{T}=k$. The singlet yield is calculated following the method in \cite{Bro76SI}. We first write the singlet fidelity as
\begin{eqnarray}
P_{S}(t)&=&\mbox{Tr}[e^{-iHt}(\rho_{0} \bigotimes_{j} \frac{\mathbb{I}_{j}}{d_{j}})e^{iHt} (|S\rangle \langle S| \bigotimes_{j}\mathbb{I}_{j})]\\
&=&\sum\limits_{m}\sum\limits_{n} \langle m|(\rho_{0}\bigotimes_{j}
\frac{\mathbb{I}_{j}}{d_{j}})|n\rangle \cdot \langle n| (|S\rangle \langle S| \bigotimes_{j} \mathbb{I}_{j})|m\rangle \cdot e^{-i(\omega_{m}-\omega_{n})t}
\end{eqnarray}
where $\rho_{0}$ is the initial state of the radical pair, and the initial state of the nuclear spins at room temperature can be approximated as $\rho _{b}(0)=\bigotimes_{j}
\mathbb{I}_{j}/d_{j}$, where $d_{j}$ is the dimension of the $j$th nuclear spin, and $\{| m\rangle \}$ and $\{| n\rangle \}$ denote the eigen states of the Hamiltonian $H$ in Eq.(1). After some calculations, we have the singlet yield as
\begin{eqnarray}
\Phi_{S}=\int_{0}^{\infty} ke^{-kt} P_{S}(t)=\frac{k}{d} \sum\limits_{m}\sum\limits_{n}  \rho_{mn} A_{nm} \frac{1}{k+i(\omega_{m}-\omega_{n})}
\end{eqnarray}
where $d=\prod d_{i}$ , $\rho_{m,n}= \langle m|( \rho_{0} \bigotimes_{j}
\mathbb{I}_{j})|n\rangle$ and $
A_{n,m}= \langle n|( |S\rangle \langle S|\bigotimes_{j}
\mathbb{I}_{j})|m\rangle$.

\vspace{1cm}

\begin{figure}[th]
\begin{center}
\begin{minipage}{14cm}
\includegraphics[width=6.5cm]{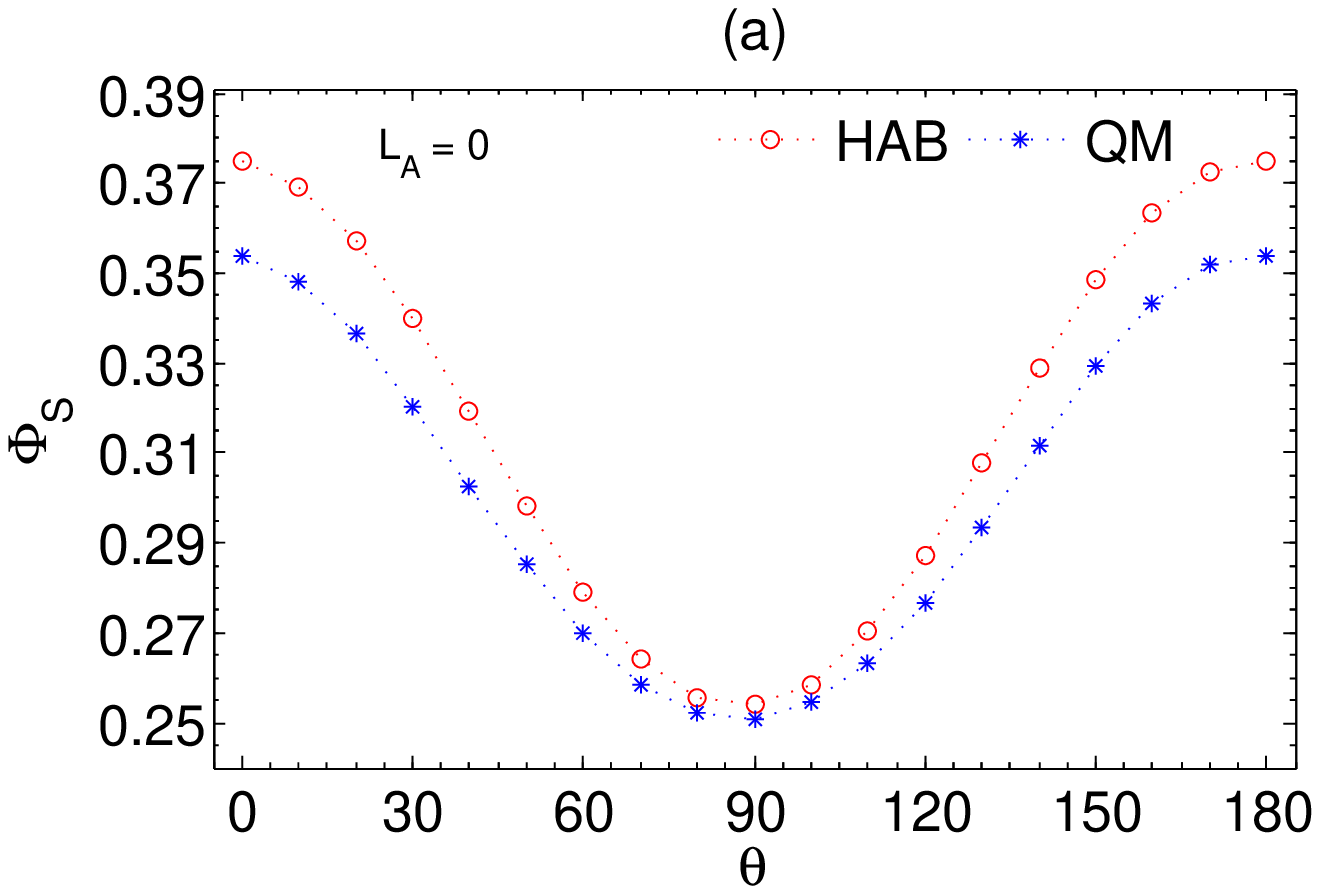}
\hspace{0.2cm}
\includegraphics[width=6.5cm]{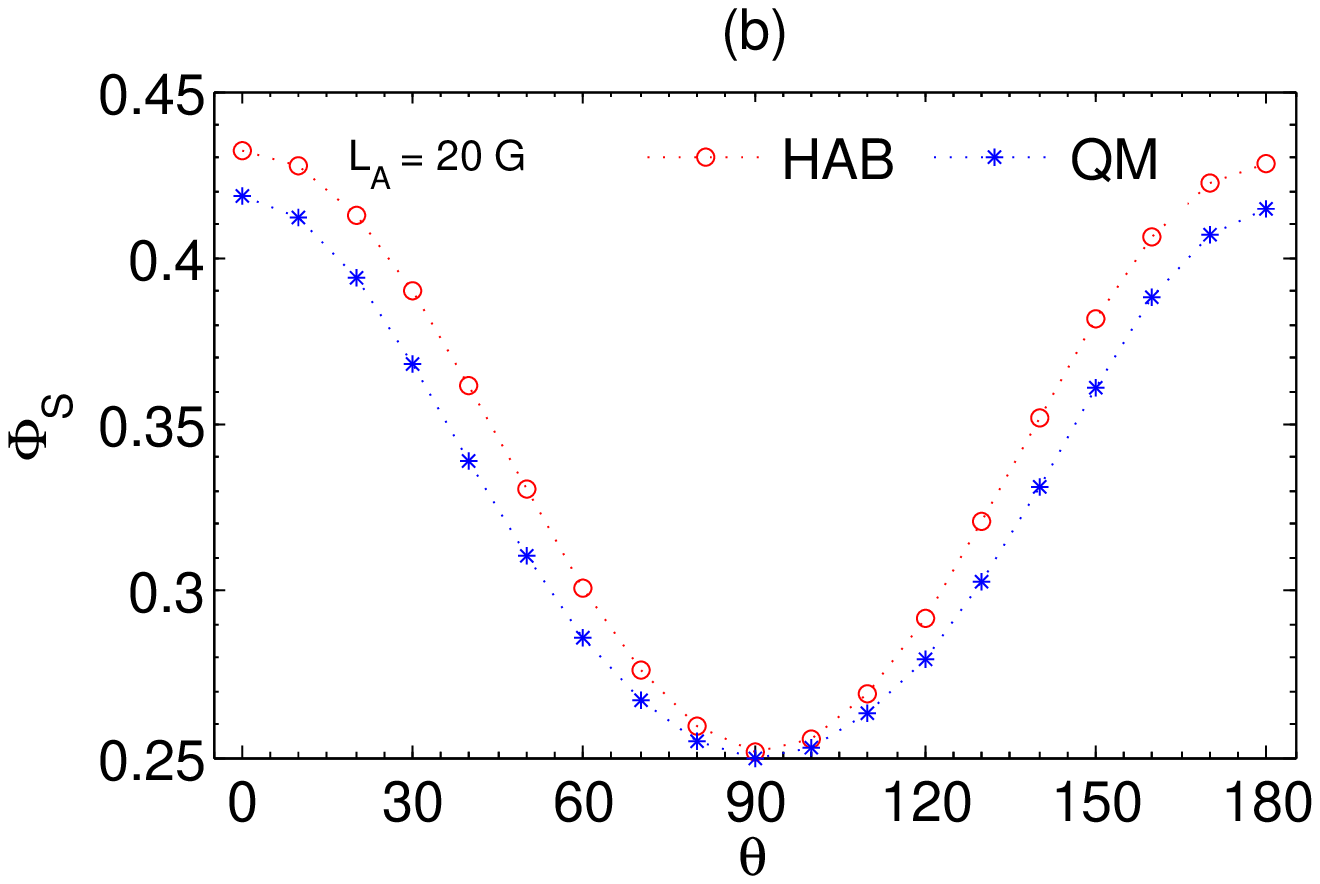}
\end{minipage}
\begin{minipage}{14cm}
\includegraphics[width=6.5cm]{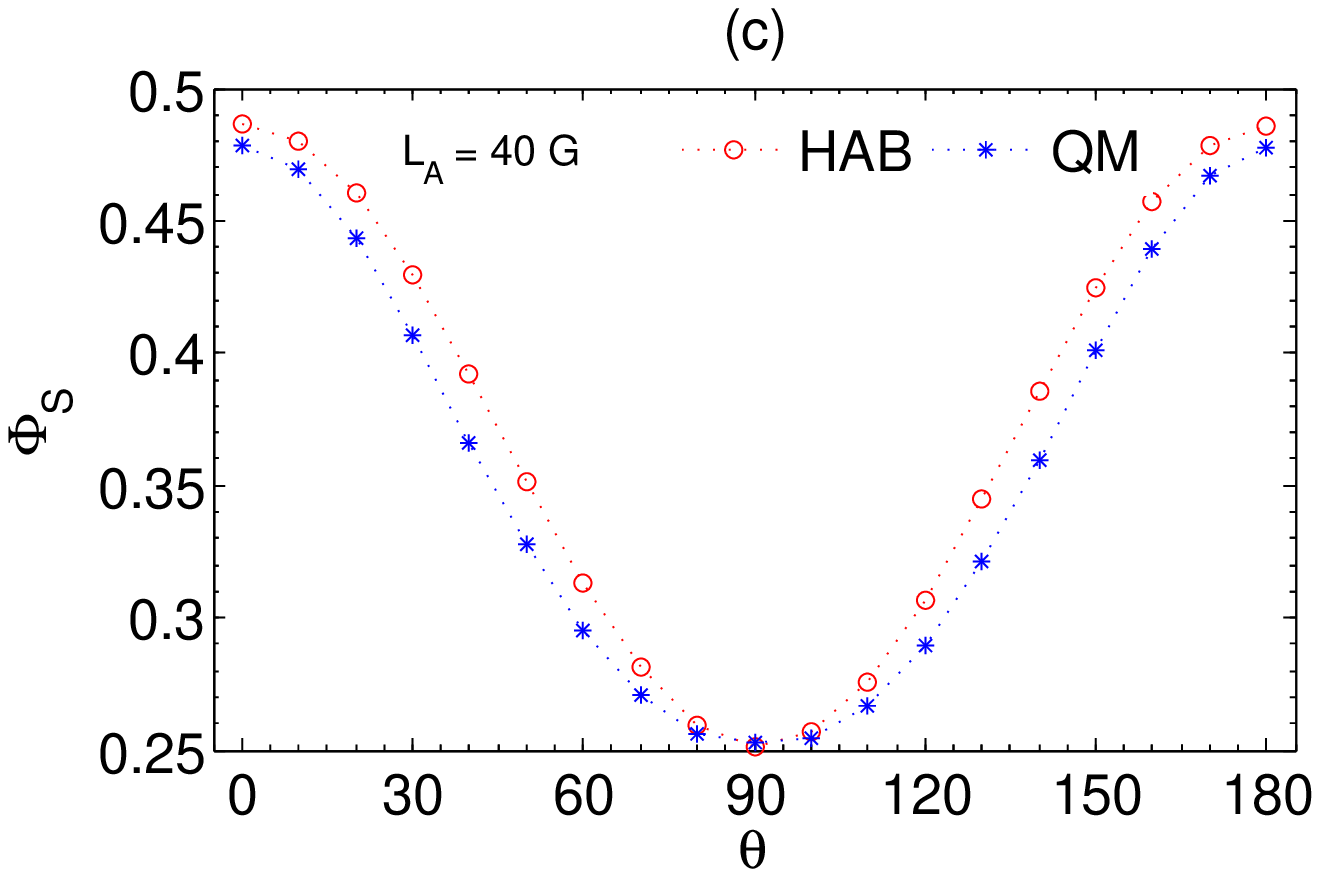}
\hspace{0.2cm}
\includegraphics[width=6.5cm]{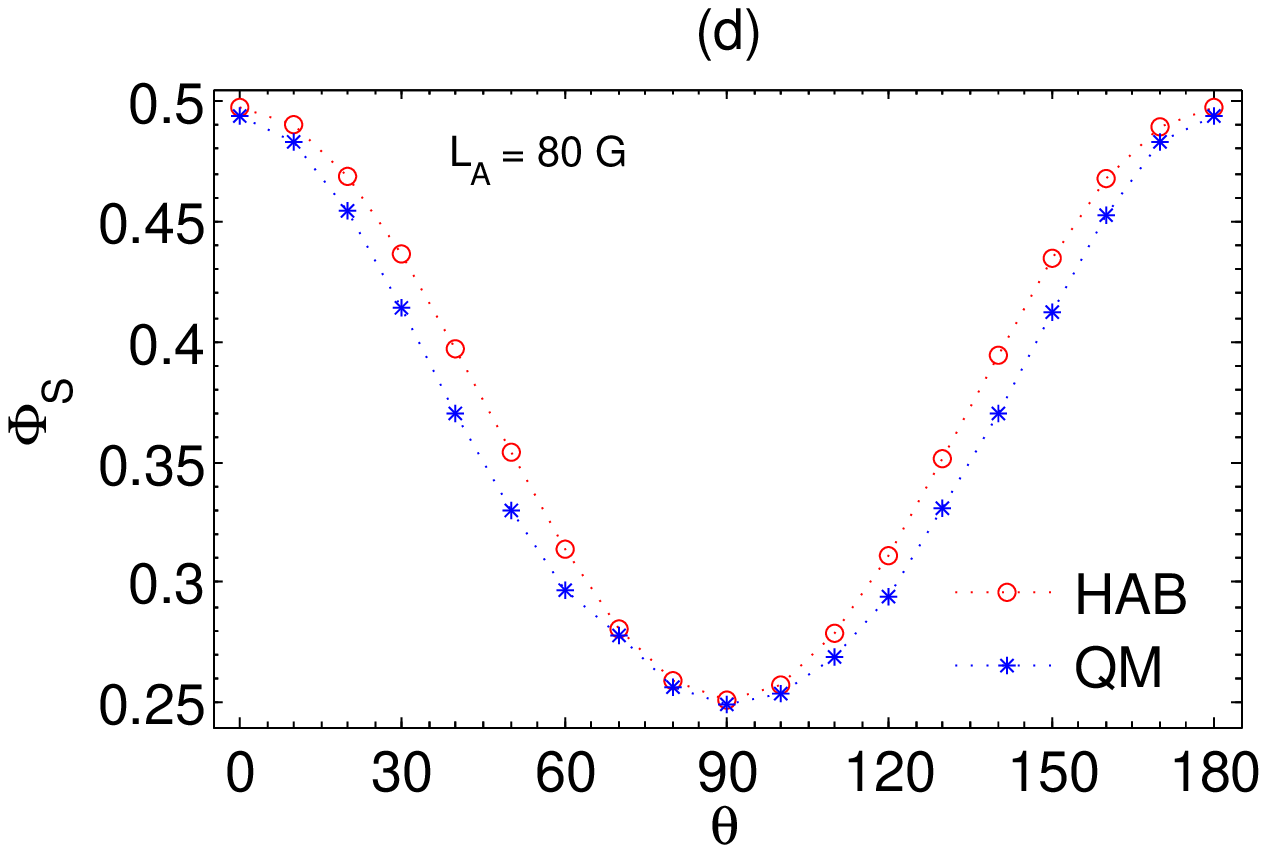}
\end{minipage}
\end{center}
\caption{(Color online) A comparison between the singlet yield from Haberkorn appraoch (HAB, red) and quantum measurement master equation (QM, blue). The radical pair life time is $\tau=1/k=2 \mu s$. The other parameters are the same as Fig.2 (a) in the main text.}\label{COMPSinYVF}
\end{figure}

\begin{figure}[bth]
\begin{center}
\begin{minipage}{14cm}
\includegraphics[width=6.5cm]{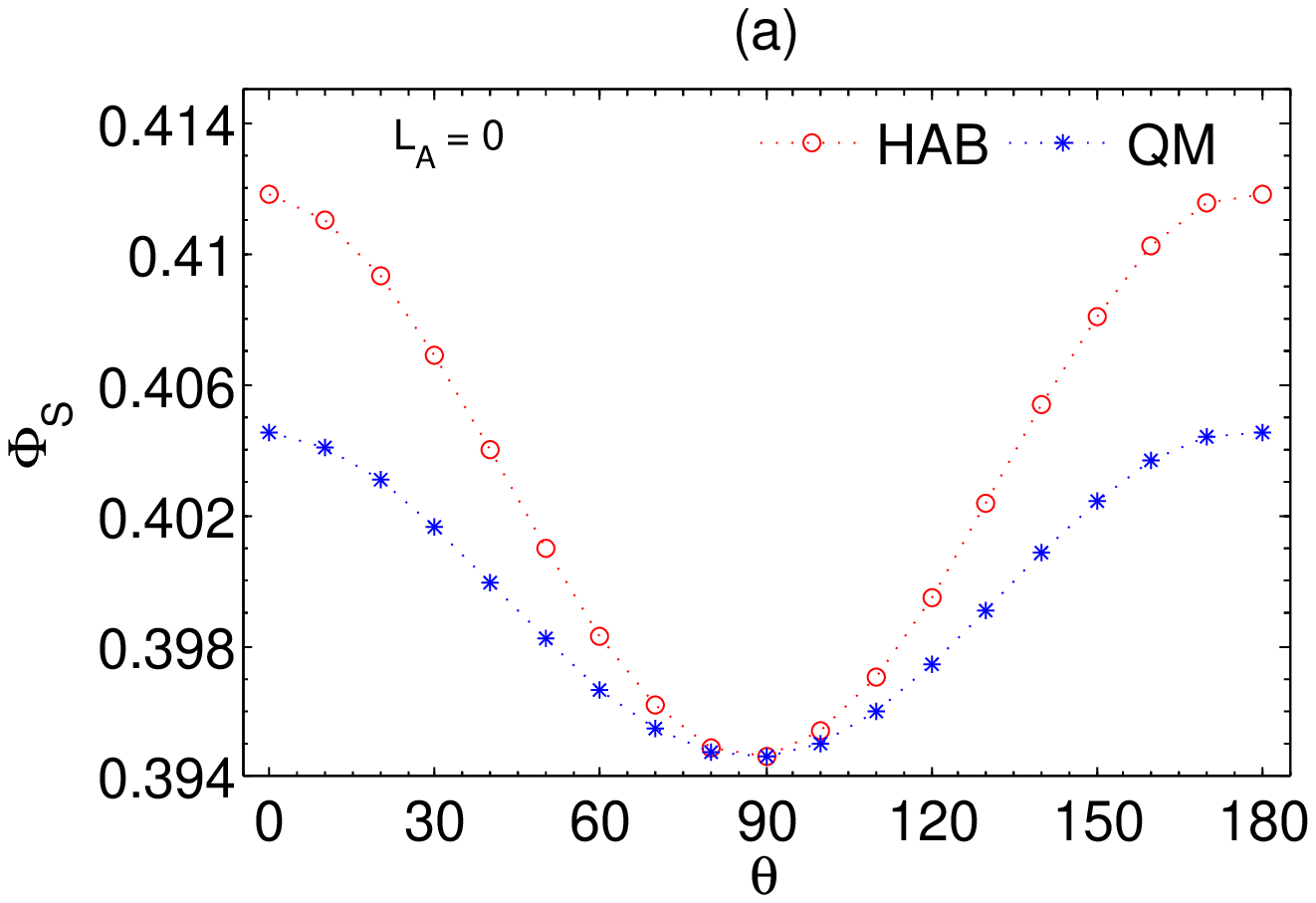}
\hspace{0.2cm}
\includegraphics[width=6.5cm]{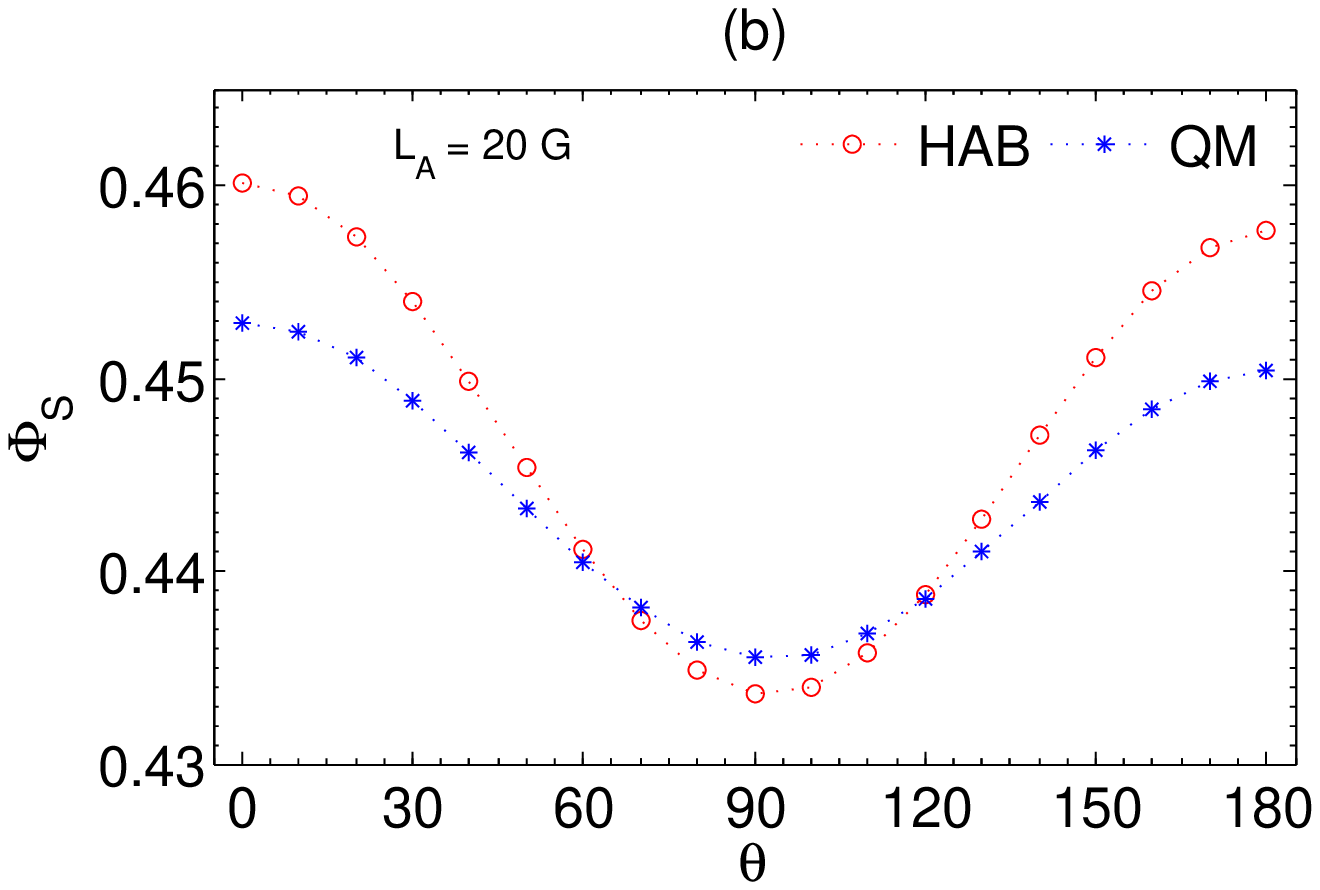}
\end{minipage}
\begin{minipage}{14cm}
\includegraphics[width=6.5cm]{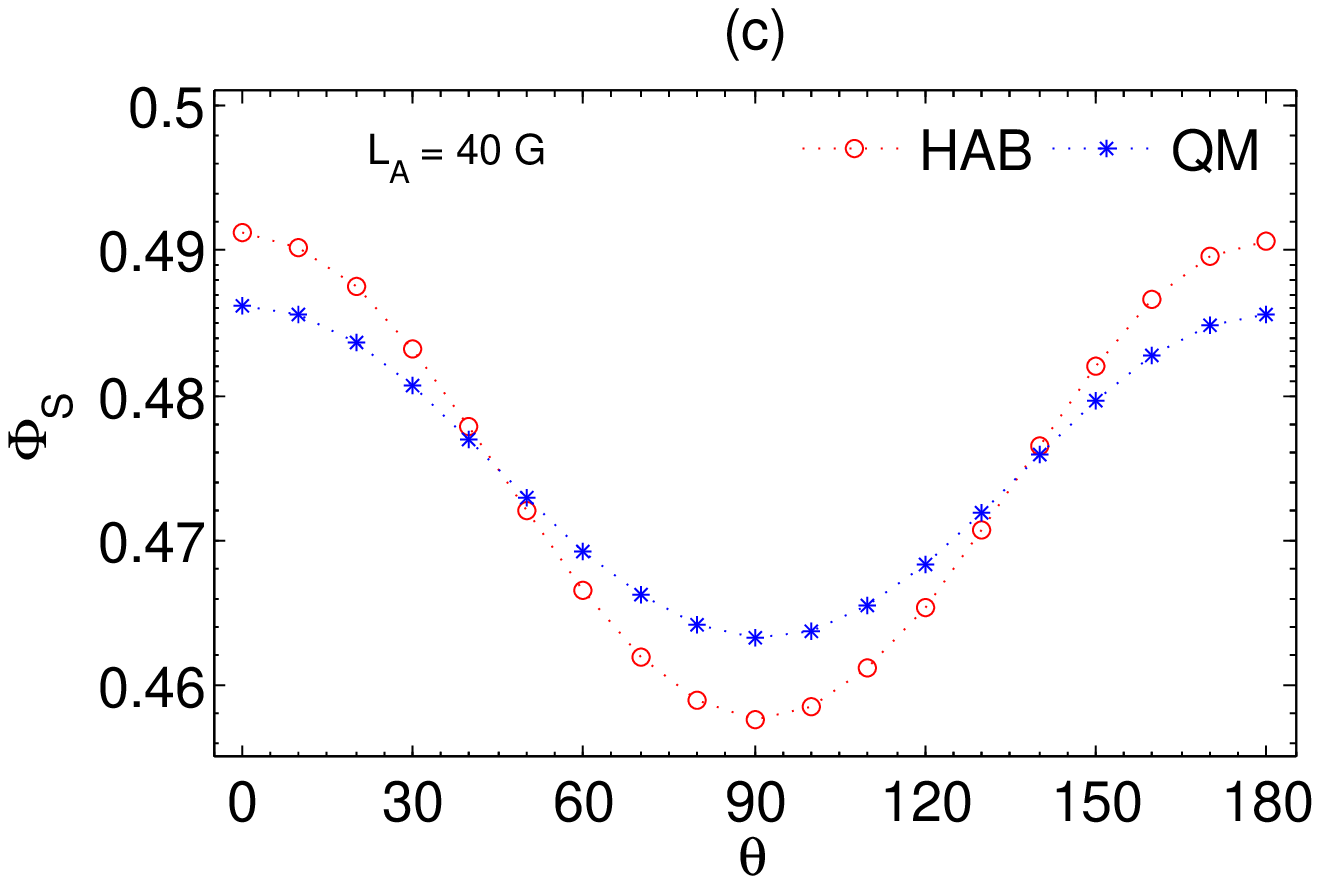}
\hspace{0.2cm}
\includegraphics[width=6.4cm]{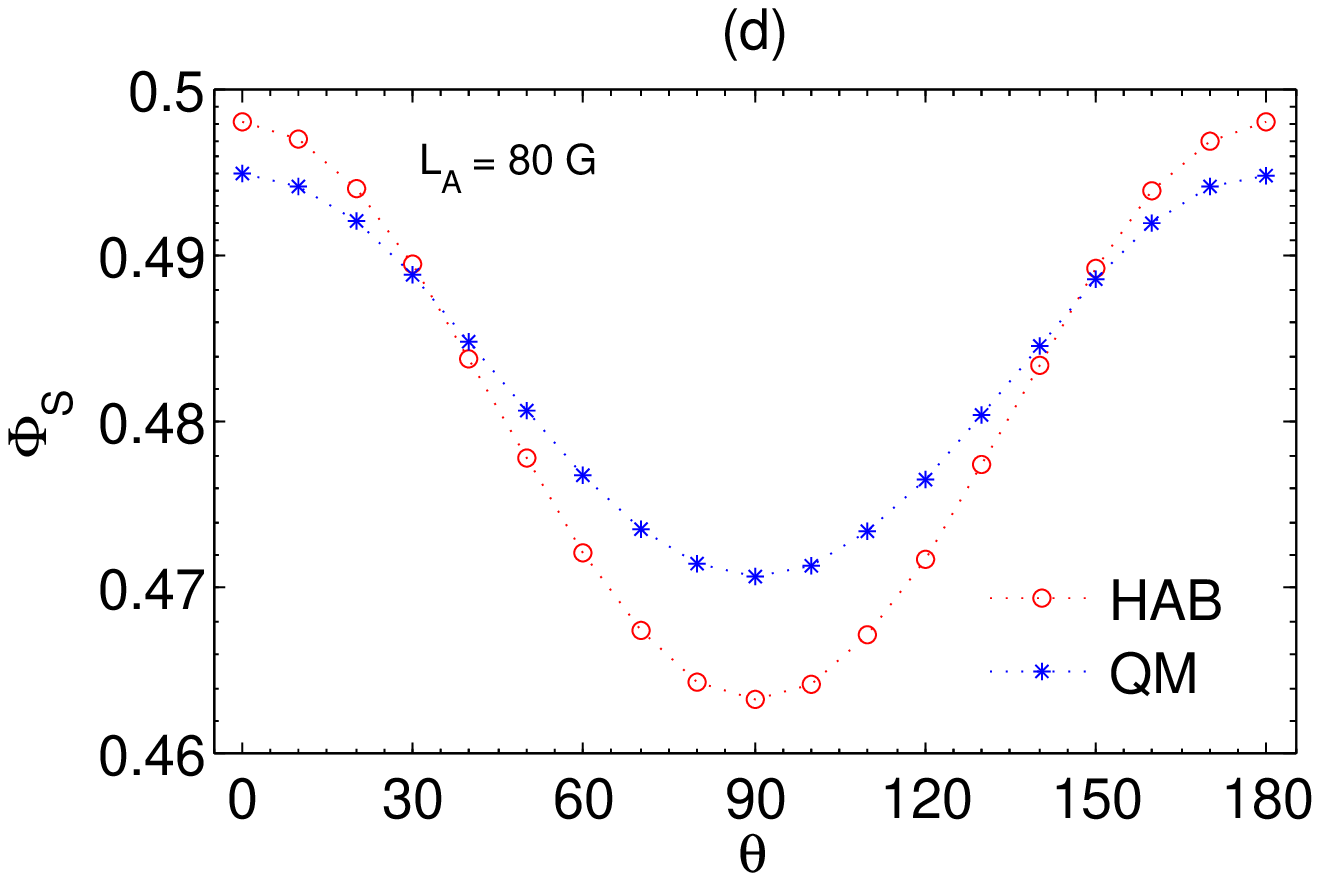}
\end{minipage}
\end{center}
\caption{(Color online) A comparison between the singlet yield from Haberkorn appraoch (HAB, red) and quantum measurement master equation (QM, blue). The radical pair life time is $\tau=1/k=50 \mbox{ns}$. The other parameters are the same as Fig.2 (a) in the main text.}\label{COMPSinYVFSL}
\end{figure}

\textsf{\bf Comparison between Haberkorn approach and quantum measurement master equation.---} In the main text, we consider the radical pair reaction with the same singlet and triplet recombination rate, i.e. $k_{S}=k_{T}=k$. The method we use to calculate the singlet yield is based on the Haberkorn approach \cite{HabSI} that describes the recombination of radical pairs
\begin{equation}
\frac{d\rho}{dt}=-i[H,\rho]-\frac{k_{S}}{2}(Q_{S}\rho+\rho Q_{S})-\frac{k_{T}}{2}(Q_{T}\rho+\rho Q_{T}) \label{HAB}
\end{equation}
where $Q_{S}$ and $Q_{T}$ are the projection operators for the singlet and triplet electronic states of the radical pair. There are alternative master equations based on quantum measurement that have been proposed to describe the recombination of radical pairs \cite{Kom09SI,JonHor10SI}. Under the condition $k_{S}=k_{T}=k$, these master equations \cite{Kom09SI,JonHor10SI} can be written in the following form
\begin{equation}
\frac{d\rho}{dt}=-i[H,\rho]-(k_{S}+k_{T})\rho+k_{S}Q_{T}\rho Q_{T}+k_{T}Q_{S}\rho Q_{S} \label{JH}
\end{equation}
From Eq.(\ref{HAB}) or Eq.(\ref{JH}), one can obtain the density matrix of the radical pair state $\rho(t)$ at time $t$ and thus calculate the singlet yield  as
\begin{equation}
\Phi_{S}=k\int_{0}^{\infty} \mbox{Tr}\left[ Q_{S}\rho(t)\right] dt
\end{equation}
We compare the results of the singlet yield from these two approaches. For the simplicity of calculation, we take the three most significant hyperfine interactions in FADH$^{.}$, i.e. those for the nitrogens N5 and N10 and the proton H5 \cite{Hore2003SI}. It can be seem from Fig.~\ref{COMPSinYVF} (for the long lifetime $\tau=1/k=2 \mu s$) and Fig.~\ref{COMPSinYVFSL} (for the short lifetime $\tau=1/k=50 \mbox{ns}$) that the difference between the results from two approaches is very small (around $1\sim 2 \%$).

\vspace{1cm}

\textsf{\bf Probe spin correlations with gradient fields.---} We assume that the gradient field on the acceptor $\vec{L}_{A} $ is much larger than the hyperfine couplings and the weak magnetic field $\vec{B}$. To calculate the singlet yield, for simplicity, the Hamiltonian can be approximated as \cite{note}
\begin{equation}
H\simeq -\gamma_{e}(\vec{L}_{A} \cdot \vec{S}_{A}+ \vec{B} \cdot \vec{S}_{D})
\end{equation}
We denote the density matrix of the initial radical pair state as $\rho$, and use the eigen states of $\vec{L}_{A} \cdot \vec{S}_{A}$ ($\{|u_{0}\rangle,|u_{1}\rangle\} $), namely
$\vec{L}_{A} \cdot \vec{S}_{A}|u_{0}\rangle=L_{A} |u_{0}\rangle $ and $\vec{L}_{A} \cdot \vec{S}_{A}|u_{1}\rangle=-L_{A} |u_{1}\rangle $, as the spin basis of the acceptor. The initial state can then be written as
\begin{equation}
 \rho=\sum\limits_{m,n=0,1}|u_{m}\rangle _{A} \langle u_{n}|\otimes \rho_{D}^{mn} \quad \mbox{where} \quad \rho_{D}^{m,n}=\langle u_{m}|\rho|u_{n}\rangle
\end{equation}
The singlet fidelity at time $t$ is
\begin{eqnarray}
P_{S}(t)&=&\langle S|\left[ |u_{0}\rangle \langle u_{0}|\otimes (U_{D} \rho^{00}_{D}U_{D}^{\dagger})\right]|S\rangle
+  \langle S|\left[ |u_{1}\rangle \langle u_{1}|\otimes( U_{D} \rho^{11}_{D}U_{D}^{\dagger}\right]|S\rangle\\\nonumber
&+&  e^{i2\gamma_{e}tL_{A}}\langle S|\left[|u_{0}\rangle \langle u_{1}|\otimes( U_{D} \rho^{01}_{D}U_{D}^{\dagger})\right]|S\rangle
+ e^{-i2\gamma_{e}t L_{A}} \langle S|\left[ |u_{1}\rangle \langle u_{0}|\otimes (U_{D} \rho^{10}_{D}U_{D}^{\dagger})\right]|S\rangle
\end{eqnarray}
where $U_{D}=\exp{(i \gamma_{e}t \vec{B}\cdot \vec{S}_{D})} $. If $L_{A}$ is very large, the last two terms (second line) in the above equation oscillate very fast and make no effective contribution to the singlet yield due to time average, thus the singlet yield will be
\begin{eqnarray}
\Phi_{S} (\vec{L}_{A},\vec{B}) &=& \int_{0}^{\infty} f(t) \cdot \{\langle S|\left[ |u_{0}\rangle \langle u_{0}|\otimes (U_{D} \rho^{00}_{D}U_{D}^{\dagger})\right]|S\rangle
+  \langle S|\left[ |u_{1}\rangle \langle u_{1}|\otimes( U_{D} \rho^{11}_{D}U_{D}^{\dagger}\right]|S\rangle\} dt\\
&=&\frac{1}{2}\int_{0}^{\infty} f(t) \cdot \mbox{Tr} ( \rho^{00}_{D}U_{D}^{\dagger}|u_{1}\rangle \langle u_{1}|U_{D}+ \rho^{11}_{D}U_{D}^{\dagger}|u_{0}\rangle \langle u_{0}|U_{D})dt\\
&=&\int_{0}^{\infty} f(t) \cdot \mbox{Tr}[(\frac{I}{4}-\frac{\hat{A}\otimes U_{D}^{\dagger} \hat{A}U_{D}}{4})\rho]dt\\
&=& \frac{1}{4}-\frac{1}{4}\langle \hat{A}\otimes \hat{V}\rangle
\end{eqnarray}
where $\hat{A}=|u_{0}\rangle\langle u_{0}|-|u_{1}\rangle \langle u_{1}|$ and $\hat{V}=\int_{0}^{\infty} dt f(t) U_{D}^{\dagger} \hat{A}U_{D}$. By choosing $\vec{L}_{A}$ in the direction of $\hat{x}$, $\hat{y}$, and $\hat{z}$, the corresponding operator $\hat{A}$ will be $\hat{X}$, $\hat{Y}$, $\hat{Z}$ respectively (which are the Pauli operators). As an example, we assume that $\hat{A}=\hat{Z}$, it can be seen that if we choose $\vec{B}$ in the direction of $\hat{x}$, $\hat{y}$, and $\hat{z}$ respectively, the corresponding operators of $\hat{V}$ are $c \hat{Z}-s  \hat{Y}$, $c \hat{Z}+s \hat{X}$, $\hat{Z}$, with $c=\int_{0}^{\infty} dt f(t) \cos{(\gamma_{e}t B)}=k^2/[k^2+(\gamma_{e}B)^{2}]$ and $s=\int_{0}^{\infty} dt f(t) \sin{(\gamma_{e}t B)}=k \gamma_{e} B/[k^2+(\gamma_{e}B)^{2}]$. These operators are linear independent. From the singlet yields corresponding to these choices of $\hat{L}_{A}$ and $\hat{B}$, we can have three independent equations, following which we can infer the spin correlations $\langle \hat{Z} \otimes \hat{X}\rangle $, $\langle \hat{Z} \otimes \hat{Y}\rangle$, $\langle \hat{Z} \otimes \hat{Z}\rangle$. In a similar way, we can choose $\hat{A}=\hat{X}, \hat{Y}$ and obtain the spin correlations
$\langle \hat{X} \otimes \hat{X}\rangle $, $\langle \hat{X} \otimes \hat{Y}\rangle$, $\langle \hat{X} \otimes \hat{Z}\rangle$, and $\langle \hat{Y} \otimes \hat{X}\rangle $, $\langle \hat{Y} \otimes \hat{Y}\rangle$, $\langle \hat{Y} \otimes \hat{Z}\rangle$.


\begin{thebibliography}{99}
\bibitem{COMP} In this case, the singlet yields calculated from different approaches are expected to be very similar \cite{Kom09,JonHor10,Maeda10,Shu10}, see also \cite{SI} for a comparison with the present model.
\end{thebibliography}
\begin{thebibliography}{99}
\bibitem{note} The approximation mainly results in a correction to the singlet fidelity in the second line of Eq.(17), which
nonetheless will not affect the singlet yield due to its fast oscillation.
\end{thebibliography}
\end{document}